\documentclass[final, journal, letterpaper]{IEEEtran}

\usepackage[utf8]{inputenc}
\usepackage{graphicx}
\usepackage{amssymb}
\usepackage{amsmath}
\usepackage{color}
\usepackage{theorem}
\usepackage{pifont}
\usepackage{array}
\usepackage{hyperref}
\usepackage{cite}
\usepackage{url}
\usepackage{pgfplots}
\usepackage{booktabs}
\usepackage[english]{babel}
\usepackage[english = american]{csquotes}
\usepackage{mathtools}
\MakeOuterQuote{"} 
\usepackage[ruled,vlined,titlenumbered]{algorithm2e}
\makeatletter
\newcommand{\removelatexerror}{\let\@latex@error\@gobble}
\makeatother

\usepackage{pgf}
\usepackage{tikz}
\usetikzlibrary{matrix,chains,positioning,arrows,calc,decorations,plotmarks,patterns, fit,backgrounds}
\usepackage{pgfplots}
\usetikzlibrary{external}                       
\tikzsetexternalprefix{tikzpic/}                   

\pgfplotsset{compat=1.3}
\tikzstyle{help lines}=[black!20,dashed]

\pgfplotscreateplotcyclelist{mylist}{red,blue,black,yellow,brown}

\pgfdeclarepatternformonly{my north east lines}{\pgfqpoint{-1pt}{-1pt}}{\pgfqpoint{8pt}{8pt}}{\pgfqpoint{6pt}{6pt}}%
{
  \pgfsetlinewidth{0.4pt}
  \pgfpathmoveto{\pgfqpoint{0pt}{0pt}}
  \pgfpathlineto{\pgfqpoint{6.1pt}{6.1pt}}
  \pgfusepath{stroke}
}
\definecolor{light_gray}{rgb}{0.6,0.6,0.6}
\definecolor{awgray}{rgb}{0.7,0.7,0.7}
\definecolor{awgray_dark}{rgb} {0.4,0.4,0.4}

\tikzset{
    >=stealth',
    mycircle/.style={circle, draw=gray, very thick, text width=.1em, minimum height=1.5em, text centered}, 
    mycircle_small/.style={circle,draw=awgray_dark,fill = awgray_dark, inner sep=0,minimum size=.6em},       
    mycircle_small_black/.style={circle,draw=black,fill = black, inner sep=0,minimum size=.6em},   
    mybox/.style={rectangle,rounded corners,draw=black, thick,text width=1em,minimum height=4em,minimum width=4em,text centered},     
    mybox_small/.style={rectangle,rounded corners,draw=black, thick,text width=1em,minimum height=2em,minimum width=2em,text centered},               
    mybox_vec/.style={rectangle,rounded corners,draw=black, thick,text width=1em,minimum height=0.7em, minimum width=4em,text centered},  
    mybox_vec_short/.style={rectangle,rounded corners,draw=black, thick,text width=1em,minimum height=0.7em, minimum width=2em,text centered},                  
    pil/.style={->, thick, shorten <=2pt, shorten >=2pt,},
}

\newtheorem{theorem}{Theorem}
\newtheorem{definition}{Definition}
\newtheorem{lemma}{Lemma}
\newtheorem{corollary}{Corollary}

\newtheorem{example}{Example}

\newcommand{\Z}[1]{\ensuremath{\mathbb{Z}_{#1}}} 
\newcommand{\Fqm}{\ensuremath{\mathbb F_{q^m}}}

\newcommand{\Fq}{\ensuremath{\mathbb F_{q}}}

\newcommand{\F}{\ensuremath{\mathbb F}}

\newcommand{\myset}[1]{\mathcal{#1}}
\newcommand{\intervallincl}[2]{\ensuremath{[#1,#2]}}
\newcommand{\intervallexcl}[2]{\ensuremath{[#1,#2-1]}}

\newcommand{\NormbasisOrdered}{\boldsymbol{\Normelement}}

\newcommand{\Normelement}{\beta}

\newcommand{\mymap}[1]{\textup{{#1}}}
\newcommand{\extsmallfield}{\ensuremath{\Phi_{\NormbasisOrdered}}}
\newcommand{\extsmallfieldInverse}{\ensuremath{\Phi^{-1}_{\NormbasisOrdered}}}
\newcommand{\extsmallfieldinput}[1]{\ensuremath{\Phi_{\NormbasisOrdered}\left(#1\right)}}
\newcommand{\extsmallfieldinputInverse}[1]{\ensuremath{\Phi^{-1}_{\NormbasisOrdered}\left(#1\right)}}
\newcommand{\liftmap}[1]{\mymap{lift}(#1)}
\newcommand{\liftmapaff}[1]{\ensuremath{\mymap{lift}_a}(#1)}
\newcommand{\OCompl}[1]{\ensuremath{\mathcal{O}({#1})}}

\SetAlgoCaptionSeparator{.} 

\newcommand{\printalgoIEEEWidth}[2]
{\begin{center}
\vspace{-2ex}
\scalebox{0.88}{
\begin{tabular}{p{#2\textwidth}}
\removelatexerror
\begin{algorithm}[H]
 #1
\end{algorithm}
\end{tabular}}
\vspace{-2ex}
\end{center}}

\DeclareMathOperator{\defi}{def}
\newcommand{\defeq}{\overset{\defi}{=}}

\DeclareMathOperator{\wt}{wt}
\DeclareMathOperator{\rank}{rank}
\DeclareMathOperator{\rk}{rk}

\DeclareMathOperator{\RRE}{RRE}

\renewcommand{\vec}[1]{\ensuremath{\mathbf{#1}}}

\newcommand{\vecelements}[1]{\ensuremath{(#1_0 \ #1_1 \ \dots \ #1_{n-1})}}
\newcommand{\vecelementsm}[1]{\ensuremath{(#1_0 \ #1_1 \ \dots \ #1_{m-1})}}

\newcommand{\MoormatExplicit}[3]{
\begin{pmatrix}
#1_{0} & #1_{1} & \dots& #1_{#3-1}\\
#1_{0}^{[1]} & #1_{1}^{[1]} & \dots& #1_{#3-1}^{[1]}\\
\vdots &\vdots&\ddots& \vdots\\
#1_{0}^{[#2-1]} & #1_{1}^{[#2-1]} & \dots& #1_{#3-1}^{[#2-1]}\\
\end{pmatrix}}

\renewcommand{\a}{\mathbf a}
\renewcommand{\b}{\mathbf b}
\renewcommand{\c}{\mathbf c}
\newcommand{\e}{\mathbf e}

\renewcommand{\r}{\mathbf r}
\newcommand{\s}{\mathbf s}
\renewcommand{\u}{\mathbf u}

\newcommand{\A}{\mathbf A}
\newcommand{\B}{\mathbf B}
\newcommand{\C}{\mathbf C}

\newcommand{\E}{\mathbf E}
\newcommand{\G}{\mathbf G}
\newcommand{\I}{\mathbf I}

\newcommand{\R}{\mathbf R}

\newcommand{\X}{\mathbf X}
\newcommand{\Y}{\mathbf Y}
\renewcommand{\Z}{\mathbf Z}
\newcommand{\0}{\mathbf 0}

\newcommand{\mycode}[1]{\ensuremath{\mathcal{#1}}}
\newcommand{\mycodeRank}[1]{\ensuremath{\mathbb{#1}}}
\newcommand{\fontmetric}[1]{\mathsf{#1}}

\newcommand{\Gab}[1]{\ensuremath{\mycode{GA}[#1]}}

\newcommand{\CDC}[1]{\ensuremath{\mycode{CD}_q(#1)}}

\newcommand{\UM}[1]{\ensuremath{\mycode{UM}(#1)}}
\newcommand{\PUM}[2]{\ensuremath{\mycode{PUM}(#1,#2)}}

\newcommand{\Subspacedist}[1]{d_{\fontmetric{S}}(#1)}
\newcommand{\SubspacedistNoInput}{d_{\fontmetric{S}}}

\newcommand{\numbRowErasures}{\varrho}
\newcommand{\numbColErasures}{\gamma}

\newcommand{\da}{\ensuremath{d_{\sigma}}}
\newcommand{\Ca}{\ensuremath{\mycode{C}_{\sigma}}}

\newcommand{\ActRowDistSet}[1]{\mathcal{C}^{(r)}_{#1}}
\newcommand{\ActColDistSet}[1]{\mathcal{C}^{(c)}_{#1}}
\newcommand{\ActRevColDistSet}[1]{\mathcal{C}^{(rc)}_{#1}}

\newcommand{\drow}[1]{\ensuremath{{d}_{#1}^{(r)}}}
\newcommand{\dcol}[1]{\ensuremath{{d}_{#1}^{(c)}}}
\newcommand{\drevcol}[1]{\ensuremath{{d}_{#1}^{(rc)}}}

\newcommand{\drowprime}[1]{\ensuremath{{d}_{#1}^{(r)\prime}}}
\newcommand{\dcolprime}[1]{\ensuremath{{d}_{#1}^{(c)\prime}}}
\newcommand{\drevcolprime}[1]{\ensuremath{{d}_{#1}^{(rc)\prime}}}

\newcommand{\drdes}[1]{\ensuremath{\delta_{#1}^{(r)}}}
\newcommand{\dcdes}[1]{\ensuremath{\delta_{#1}^{(c)}}}
\newcommand{\drcdes}[1]{\ensuremath{\delta_{#1}^{(rc)}}}

\newcommand{\drdesprime}[1]{\ensuremath{\delta_{#1}^{(r)\prime}}}
\newcommand{\dcdesprime}[1]{\ensuremath{\delta_{#1}^{(c)\prime}}}
\newcommand{\drcdesprime}[1]{\ensuremath{\delta_{#1}^{(rc)\prime}}}

\newcommand{\BMD}[1]{\ensuremath{\mathsf{BMD(#1)}}}

\newcommand{\memory}{\mu}
\newcommand{\memoryH}{\mu_H}
\newcommand{\constraintlength}{\nu}
\newcommand{\constraintlengthi}{\nu_i}

\newcommand{\wtsum}{\wt_{\Sigma}}
\newcommand{\distsum}{d_{\Sigma}}
\newcommand{\dfree}{d_{f}}

\newcommand{\slope}{\sigma}

\newcommand{\pumforward}[1]{\ell_f^{(#1)}}
\newcommand{\pumbackward}[1]{\ell_b^{(#1)}}

\newcommand{\pumforwardprime}[1]{\ell_f^{(#1)\prime}}
\newcommand{\pumbackwardprime}[1]{\ell_b^{(#1)\prime}}

\newcommand{\pumforwardBig}[1]{L_f^{(#1)}}
\newcommand{\pumbackwardBig}[1]{L_b^{(#1)}}

\newcommand{\myspace}[1]{\mathcal{#1}}
\newcommand{\Rowspace}[1]{\myspace{R}_q\left(#1\right)}
\newcommand{\RowspaceNoInput}{\myspace{R}_q}

\newcommand{\Grassm}[1]{\myspace{G}_q(#1)}

\newcommand{\quadbinom}[2]{\ensuremath{
{#1
\brack
#2}
}}

\begin{document}

\title{Convolutional Codes in Rank Metric with Application to Random Network Coding}

\IEEEoverridecommandlockouts

\author{{Antonia Wachter-Zeh, 
Markus Stinner, 
Vladimir Sidorenko 
\thanks{Parts of 
this work were presented at the \textit{IEEE International Symposium on Network Coding 2012 (NETCOD)}, Cambridge, MA, USA, 2012 \cite{WachterSidorenko-RankMetricConvolutionalNetworkCoding_conf}. 
A. Wachter-Zeh's work was supported in part by a Minerva Postdoctoral Fellowship and in part by the German Research Council "Deutsche
Forschungsgemeinschaft" (DFG) under Grant No. Bo867/21. 
M. Stinner's work was supported by an Alexander von Humboldt Professorship endowed by the German Federal Ministry of Education and Research.
V. Sidorenko's work was supported by the German Research Council (DFG) under Grant No. Bo867/22.

A. Wachter-Zeh is with the Computer Science Department, Technion---Israel Institute of Technology, Haifa, Israel (e-mail: antonia@cs.technion.ac.il).

M. Stinner is with the Institute for Communications Engineering, Technical University of Munich, Germany (e-mail: markus.stinner@tum.de).

V. Sidorenko is with the Institute for Communications Engineering, Technical University of Munich, Germany (e-mail:vladimir.sidorenko@tum.de).
}}}

\maketitle

\begin{abstract}
Random network coding recently attracts attention as a technique to disseminate information in a network.
This paper considers a non-coherent multi-shot network, where the unknown and time-variant network is used several times.
In order to create dependencies between the different shots, particular convolutional codes in rank metric are used.
These codes are so-called \emph{(partial) unit memory} ((P)UM) codes, i.e., convolutional codes with memory one. First, distance measures for convolutional codes in rank metric are shown and two constructions of (P)UM codes in rank metric based on the generator matrices of maximum rank distance codes are presented. Second, an efficient error-erasure decoding algorithm for these codes is presented. Its guaranteed decoding radius is derived and its complexity is bounded. Finally, it is shown how to apply these codes for error correction in random linear and affine network coding.
\end{abstract}

\begin{keywords}
convolutional codes, network coding, (partial) unit memory codes, rank-metric codes
\end{keywords}

\section{Introduction}\label{sec:introduction}
\emph{Random linear network coding} (RLNC, see e.g., \cite{Ahlswede_NetworkInformationFlow_2000,HoKoetterMedardKargerEffros-BenefitsCodingRouting_2003,MedardKoetterKargerEffrosShiLeong-RLNCMulticast_2006}) and more recently, \emph{random affine network coding} (RANC, see \cite{Gadouleau2011Matroid}), are powerful means for distributing information in networks. 
In these models, it is assumed that the packets are vectors over a finite field and that each internal node of the network performs a random linear (or random affine, respectively) combination of all packets received so far and forwards this random combination to adjacent nodes. Notice that affine combinations are particular linear combinations, see \cite{Gadouleau2011Matroid}.

When we consider the transmitted packets as rows of a matrix, then the linear combinations performed by the nodes are elementary row operations on this matrix.
During an error- and erasure-free transmission over such a network, the row space of the transmitted matrix is therefore preserved.
However, due to the linear combinations at the nodes, 
a single erroneous packet can propagate widely throughout the network. This makes error-correcting techniques in random networks essential.

Based on these observations, K\"otter and Kschischang \cite{koetter_kschischang} used \emph{subspace codes} for error control in RLNC and introduced a channel model, called the \emph{operator channel}.
Silva, Kschischang and K\"otter \cite{silva_rank_metric_approach} showed that lifted rank-metric block codes result in almost optimal subspace codes 
for RLNC. In particular, they used \emph{Gabidulin codes} \cite{Delsarte_1978,Gabidulin_TheoryOfCodes_1985,Roth_RankCodes_1991}, which are rank-metric analogs to Reed--Solomon codes.
Both approaches were extended to affine subspace codes by Gadouleau and Yan for error correction in RANC \cite{Gadouleau2011Matroid}.

In this paper, we consider \emph{non-coherent multi-shot} network coding (see e.g., \cite{NobregaUchoa-2010Multishot}). Therefore, we use the network several times, where the 
internal structure of the network is unknown and might change in each shot. 
Creating dependencies between the transmitted words of the different 
shots can help to cope with difficult error patterns and strongly varying channels. 
We achieve these dependencies by using \emph{convolutional} network codes. 

In particular, we {consider} so-called \emph{(partial) unit memory} ((P)UM) codes \cite{Lee_UnitMemory_1976,Lauer_PUM_1979} in rank metric. 
(P)UM codes are a special class of convolutional codes with memory one. 
They can be constructed
based on block codes, e.g., Reed--Solomon \cite{zyablov_sidorenko_pum,Pollara_FiniteStateCodes,Justesen_BDDecodingUM} or cyclic codes \cite{DettmarSorger_PUMBCH, DettmarShav_NewUMCodes}. 
The underlying block codes make an algebraic description of the convolutional code possible, enable us to estimate the distance properties and allow us to take into account existing efficient block decoders in order to decode the convolutional code. Notice that a convolutional code with arbitrary memory can be considered as PUM convolutional code with larger block size. This is another motivation to start working on (P)UM convolutional codes in rank metric; their generalization to multi-memory codes in rank metric is an interesting topic for future work.

A convolutional code in \emph{Hamming metric} can be characterized by its \emph{active row distance}, which in turn is basically determined by the \emph{free distance} and the \emph{slope}. These distance measures determine the error-correcting capability of the convolutional code. In \cite{Lee_UnitMemory_1976,Lauer_PUM_1979,Thommesen_Justesen_BoundsUM,Pollara_FiniteStateCodes}, upper bounds on the free (Hamming) distance and the slope of (P)UM codes were derived. 

In \cite{WachterSidBossZyb-PUMGabidulin_ISIT2011_conf,WachterSidBossZyb_PUMbasedGab}, distance measures for convolutional codes in rank metric were introduced and PUM codes based on the parity-check matrix of Gabidulin codes were constructed.

In this paper, we construct (P)UM codes based on the \emph{generator} matrix of Gabidulin codes and
calculate their distance properties. 
As a distance measure, the \emph{sum rank metric} is used, which is motivated by multi-shot network coding \cite{NobregaUchoa-2010Multishot} and which is used to define the \emph{free rank distance} and the \emph{active row rank distance}, see also \cite{WachterSidBossZyb-PUMGabidulin_ISIT2011_conf,WachterSidBossZyb_PUMbasedGab}. 
Moreover, we provide an efficient decoding algorithm based on rank-metric 
block decoders, which is able to handle errors and at the same time column and row erasures. This decoding algorithm can be seen as a generalization of the 
Dettmar--Sorger algorithm \cite{DettmarSorger_BMDofUM} to error-erasure decoding as well as to the rank metric. 
Further, we show how lifted PUM codes can be applied for error-correction in RLNC and RANC.

There are other contributions devoted to convolutional network codes (see e.g. \cite{Erez2004Convolutional,Li2006Convolutional,Prasad2010Networkerror,Guo2011Localized}).
However, in most of these papers, convolutional codes are used to solve the problem of efficiently mixing information in a multicast setup and none of these code constructions is based on codes in the rank metric and deals with the transmission over the operator channel as ours. 
Our contribution can be seen as an equivalent to the block code construction from \cite{silva_rank_metric_approach}.

This paper is structured as follows. 
In Section~\ref{sec:preliminaries}, definitions and notations for (lifted) rank-metric codes as well as for convolutional codes are given. 
Section~\ref{sec:conv_metrics} shows distance measures for convolutional codes in rank metric and in Section~\ref{sec:pum_constructions}, we provide two explicit constructions of (P)UM codes based on Gabidulin codes and we derive their distance properties. 
The first construction yields codes of low code rate, and is generalized by the second construction to arbitrary code rates.
In Section~\ref{sec:pum_decoding}, we present an efficient decoding algorithm based on rank-metric block decoders, which is able to handle errors and row/column erasures. This decoding algorithm can be seen as a generalization of the Dettmar--Sorger algorithm~\cite{DettmarSorger_BMDofUM}. 
In Section~\ref{sec:pum_networkcoding}, we show---similar to \cite{silva_rank_metric_approach}---how lifted (P)UM codes can be applied in RLNC and how decoding in RLNC reduces to error-erasure decoding of our (P)UM code construction, which can efficiently be decoded by our algorithm from Section~\ref{sec:pum_decoding}. Finally, Section~\ref{sec:affine_networkcoding} outlines how to apply our codes for RANC and Section~\ref{sec:conclusion} concludes this paper.

\section{Preliminaries}\label{sec:preliminaries}
\subsection{Notations}

Let $q$ be a power of a prime and let us denote the $q$-power for any positive integer $i$ by $x^{[i]}\defeq x^{q^i}$. Let $\Fq$ denote the finite field of order $q$ and $\F =\Fqm$ its extension field of order $q^m$. 
We use $\Fq^{s \times n}$ to denote the set of all $s\times n$ matrices over $\Fq$ and 
$\F^n =\F^{1 \times n}$ for the set of all row vectors of length $n$ over $\F$. 
Let $\Rowspace{\A}$ denote the row space of a matrix $\A$ over $\Fq$ and let 
$\mathbf I_{s}$ denote the $s \times s$ identity matrix.
Moreover, denote the elements of a vector $\a^{(i)} \in \F^n$ by $\a^{(i)} = \vecelements{a^{(i)}}$.
Throughout this contribution, let the rows and columns of an $m\times n$-matrix $\mathbf A$ be indexed by $0,\dots, m-1$ and $0,\dots, n-1$ and denote the set of integers $\intervallincl{a}{b} = \{i: a \leq i \leq b, i \in \mathbb{Z}\}$.

Let $\NormbasisOrdered= \vecelementsm{\beta}$ be an ordered basis of $\F$ over $\Fq$. There is a bijective map $\extsmallfield$ of any vector $\mathbf a \in \F^n$ on a matrix $\mathbf A \in \Fq^{m \times n}$, denoted as follows:
\begin{align*}
\extsmallfield:\quad\F^{n} &\rightarrow \Fq^{m \times n}\label{eq:mapping_smallfield}\\
\a = \vecelements{a} &\mapsto \A, 
\end{align*}
where $\A =\extsmallfieldinput{\vec{a}}\in \Fq^{m \times n}$ is defined such that
$a_j = \sum_{i=0}^{m-1} A_{i,j} \Normelement_i, \quad \forall j \in \intervallexcl{0}{n}$.
In the following, we use both representations (as a matrix over $\Fq$ or as a vector over $\F$), depending on what is more useful in the context. 

Consider the vector space $\Fq^n$ of dimension $n$ over $\Fq$. 
The \emph{Grassmannian} of dimension $r\leq n$ is the set of all subspaces of $\Fq^n$ of dimension $r $ and is denoted by $\Grassm{n,r}$. 
The cardinality of $\Grassm{n,r}$ is the so-called Gaussian binomial, calculated by
\begin{equation*}
\big|\Grassm{n,r}\big|=\quadbinom{n}{r} \defeq \prod\limits_{i=0}^{r-1} \frac{q^n-q^i}{q^r-q^i},
\end{equation*}
with the upper and lower bounds (see e.g. \cite[Lemma~4]{koetter_kschischang})
\begin{equation}\label{eq:bounds_gaussian_binomial}
q^{r(n-r)}\leq \quadbinom{n}{r} \leq 4 q^{r(n-r)}.
\end{equation}
For two subspaces $\myspace{U},\myspace{V}$ in {$\Fq^n$}, 
 we denote by $\myspace{U}+\myspace{V}$ the smallest subspace containing the union of $\myspace{U}$ and $\myspace{V}$. 
The \emph{subspace distance} between $\myspace{U},\myspace{V}$ in {$\Fq^n$}
 is defined by
\begin{align*}
\Subspacedist{\myspace{U},\myspace{V}}&=\dim(\myspace{U}+\myspace{V})-\dim(\myspace{U}\cap \myspace{V})\\
&=2 \dim(\myspace{U}+\myspace{V})-\dim(\myspace{U})-\dim(\myspace{V}).
\end{align*}
It can be shown that the subspace distance is indeed a metric (see e.g. \cite{koetter_kschischang}).

\subsection{Rank Metric and Gabidulin Codes}
We define the \textit{rank norm} $\rk(\a)$ as the rank of $\A  = \extsmallfieldinput{\vec{a}}\in \Fq^{m \times n}$ over $\mathbb{F}_{q}$. 
The rank distance between \vec{a} and \vec{b} is the rank of the difference of the two matrix representations (see \cite{Gabidulin_TheoryOfCodes_1985}):
\begin{equation*}
d_{\fontmetric{R}}(\a,\vec{b})\defeq \rk(\a-\vec{b}) = \rank(\A-\B).
\end{equation*}
The minimum rank distance $d$ of a {code $\mycodeRank{C} \subseteq \F^{n}$} is defined by
\begin{equation*}
d \defeq \min_{\substack{{\a,\b} \in \mycodeRank{C}\\ \a \neq \b}}
\big\lbrace d_{\fontmetric{R}}(\a,\vec{b}) = \rk(\a-\b) \big\rbrace. 
\end{equation*}
For linear codes of length $n \leq m$ and dimension $k$, the Singleton-like upper bound \cite{Delsarte_1978,Gabidulin_TheoryOfCodes_1985,Roth_RankCodes_1991} implies that $d \leq n-k+1$.
If $d=n-k+1$, the code is called a \emph{maximum rank distance} (MRD) code.

Gabidulin codes are a special class of rank-metric codes and can be defined in vector representation by its generator matrix as follows.

\begin{definition}[Gabidulin Code \cite{Gabidulin_TheoryOfCodes_1985}]
A linear $\Gab{n,k}$ code $\mycodeRank{C} \subseteq \F^n$ of length $n \leq m$ 
and dimension $k$ is defined by its $k \times n$ generator matrix $\mathbf G_{\mycode{G}}$:
\begin{equation*}
\mathbf G_{\mycode{G}} = 
\MoormatExplicit{g}{k}{n},
\end{equation*}
where $ g_0, g_1, \dots, g_{n-1} \in \F$ are linearly independent over $\Fq$. 
\end{definition}
Gabidulin codes are MRD codes, i.e., $d=n-k+1$, see \cite{Gabidulin_TheoryOfCodes_1985}.

Let the matrix $\C\in \Fq^{m \times n} = \extsmallfieldinput{\c}$, where $\mathbf c = \u \cdot \G_{\mycode{G}}$ for some $\u \in\F^k$, be a transmitted codeword that is corrupted by an additive error matrix $\mathbf E \in \Fq^{m \times n}$. At the receiver side, only the received matrix $\mathbf R \in \Fq^{m \times n}$, where $\mathbf R = \mathbf C+ \mathbf E$, is known. 
The channel might provide additional side information in the form of erasures, which help to increase the decoding performance.
This additional side information of the channel is assumed to be given in form of:
\begin{itemize}
\item $\numbRowErasures$ {row erasures} (in \cite{silva_rank_metric_approach} called "deviations") and
\item $\numbColErasures$ {column erasures} (in \cite{silva_rank_metric_approach} called "erasures"),
\end{itemize}  
such that the received matrix can be decomposed into 
\begin{equation}\label{eq:decomp_errrorerasures}
\R
=\C+ \underbrace{\A^{(R)} \B^{(R)} + \A^{(C)} \B^{(C)} + \A^{(E)} \B^{(E)}}_{= \E},
\end{equation}
where $\A^{(R)} \in \Fq^{m \times \numbRowErasures}$, $\B^{(R)} \in \Fq^{\numbRowErasures \times n}$, 
$\A^{(C)} \in \Fq^{m \times \numbColErasures}$, $\B^{(C)} \in \Fq^{\numbColErasures\times n}$,
$\A^{(E)} \in \Fq^{m \times t}$, $\B^{(E)} \in \Fq^{t \times n}$ { are full rank matrices}.
The channel outputs $\R$ and additionally $\A^{(R)}$ and $\B^{(C)}$ to the receiver. Further, $t$ denotes the number of errors without side information. 
The decomposition from \eqref{eq:decomp_errrorerasures} is not necessarily unique, but we can use any of them.

The rank-metric block \emph{bounded minimum distance} (BMD) error-erasure decoding algorithms from \cite{GabidulinPilipchuck_ErrorErasureRankCodes_2008,silva_rank_metric_approach}
can reconstruct any $\c \in \Gab{n,k}$ from $\r =\extsmallfieldinputInverse{\R}$ with complexity $\mathcal O(n^2)$ operations over $\F$ if 
\begin{equation}\label{eq:errorerasurecond}
2t + \numbRowErasures + \numbColErasures \leq d-1 = n-k.
\end{equation}

\subsection{Lifted Gabidulin Codes}
A \emph{constant-dimension code} is a subset of a certain Grassmannian.
We shortly recall the definition from \cite{silva_rank_metric_approach} of a special class of {constant-dimension codes}, called \emph{lifted Gabidulin codes}.

Let $\CDC{n,M_{\fontmetric{S}},\SubspacedistNoInput,r}$ denote a constant-dimension code in $\Grassm{n,r}$ with cardinality $M_{\fontmetric{S}}$ and minimum subspace distance $\SubspacedistNoInput$. 
The \emph{lifting} of a block code is defined as follows.
\begin{definition}[Lifting of Matrix or Code]\label{def:lifting_matrixcode}
Consider the map
\begin{align*}
\mymap{lift}:\quad \Fq^{r \times (n-r)} &\rightarrow \Grassm{n,r}\\
\X &\mapsto \Rowspace{[\I_r \ \X]}.
\end{align*}
The subspace $\liftmap{\X} = \Rowspace{[\I_r \ \X]}$ is called \textbf{lifting} of the matrix $\X$. If we apply this map on all codewords (in matrix representation) of a code $\mycodeRank{C}$, then the constant-dimension code $\liftmap{\mycodeRank{C}}$ is called lifting of $\mycodeRank{C}$.
\end{definition}

The following lemma shows the properties of a lifted Gabidulin code.
\begin{lemma}[Lifted Gabidulin Code \cite{silva_rank_metric_approach}]\label{lem:subspacecode_lifted_mrd}
Let $\mycodeRank{C}$ be a Gabidulin $\Gab{r,k}$ code over $\mathbb{F}_{q^{n-r}}$ of length $r\leq n-r$, minimum rank distance $d=r-k+1$ and cardinality $M_{\fontmetric{R}}=q^{(n-r)k}$.
  
Then, the lifting of the transposed codewords, i.e.,
\begin{equation*}
\liftmap{\mycodeRank{C}^T}\defeq\Big\{\liftmap{\C^T} = \Rowspace{[\I_r \ \C^T]} : \extsmallfieldinputInverse{\C} \in \mycodeRank{C}\Big\}
\end{equation*}
is a $\CDC{n,M_{\fontmetric{S}},\SubspacedistNoInput,r}$ constant-dimension code of cardinality $M_{\fontmetric{S}}=M_{\fontmetric{R}}=q^{(n-r)k}$, minimum subspace distance $\SubspacedistNoInput=2d$ and lies in the Grassmannian $\Grassm{n,r}$.
\end{lemma}

\subsection{Convolutional Codes and (Partial) Unit Memory Codes}
In practical realizations, it does not make sense to consider (semi-)infinite sequences and therefore, we consider only linear \emph{zero-forced terminated} convolutional codes. 
Such a convolutional code $\mycode{C}$ is defined
by the following  terminated {non-catastrophic} $N k \times (n(N+\memory))$ generator matrix $\G$ over $\F$, for some integer $N$: 
\begin{equation}\label{eq:genmat_termin}
\G = 
{ 
{\setlength\arraycolsep{0.4em}
\left(\begin{array}{ccccccc}
\G^{(0)} & \G^{(1)} & \dots & \G^{(\memory)} && \\
&\G^{(0)} & \G^{(1)} & \dots & \G^{(\memory)}&& \\
&\quad \ddots&\ddots&\ddots&\ddots&\\
&&\G^{(0)} & \G^{(1)} & \dots & \G^{(\memory)}\!\!\!\!\!\!\!
\end{array}\right),
}}
\end{equation}
where $\mathbf G^{(i)}$, $\forall i=0,\dots,\memory$, are $k \times n$-matrices and 
$\memory$ denotes the \emph{memory} of $\G$, see~\cite{Johannesson_Fund_Conv_Codes} and Definition~\ref{def:memory}.
Each codeword of $\mycode{C}$ is a sequence of $N+\memory$ blocks of length $n$ over $\F$, i.e., 
$\c = (\c^{(0)}\ \c^{(1)}\  \dots \ \c^{(N +\memory-1)})$, represented equivalently as a sequence of $m\times n$ matrices over $\Fq$, i.e., $\C = \extsmallfieldinput{\c} =  (\C^{(0)}\ \C^{(1)}\  \dots \ \C^{(N +\memory-1)})$.

Memory and constraint length are properties of the generator matrix. 
We follow Forney's notations \cite{Forney_ConvolutionalCodesAlg} based on the polynomial representation of the generator matrix:
\begin{align*}
\mathbf G (D) &= \G^{(0)} + \G^{(1)} D + \G^{(2)} D^2 + \dots+\G^{(\memory)} D^\memory\\
&=\big(g_{i,j}(D)\big)^{i \in \intervallexcl{0}{k}}_{j \in \intervallexcl{0}{n}},
\end{align*}
where
$g_{i,j}(D) = g_{i,j}^{(0)} + g_{i,j}^{(1)}D  +\dots+ g_{i,j}^{(\memory)}D^\memory$ and $g_{i,j}^{(l)} \in \Fq$, $\forall l\in\intervallincl{0}{\memory}$,
$\forall i\in\intervallexcl{0}{k}$ and $\forall j\in\intervallexcl{0}{n}$.
\begin{definition}[Constraint Length and Memory]\label{def:memory}
The $i$-th constraint length $\constraintlengthi$ of a polynomial generator matrix $\mathbf G(D)$ is
\begin{equation*}
\constraintlengthi \defeq \max_{\substack{j\in\intervallexcl{0}{n}}} \big\lbrace \deg g_{i,j}(D) \big\rbrace, \quad \forall i \in \intervallexcl{0}{k}.
\end{equation*}
The memory of $\mathbf G(D)$ is 
\begin{equation*}
\memory \defeq \max_{\substack{i\in\intervallexcl{0}{k}}} \lbrace \constraintlengthi \rbrace,
\end{equation*}
and the overall constraint length of $\mathbf G(D)$ is 
$\constraintlength \defeq \sum_{i=0}^{k-1} \nu_i$.
\end{definition}

(P)UM codes are a special class of convolutional codes of memory $\memory=1$, introduced by Lee and Lauer \cite{Lee_UnitMemory_1976,Lauer_PUM_1979}. The semi-infinite generator matrix consists therefore
of two $k\times n$ submatrices $\G^{(0)}$ and $\G^{(1)}$. 
These matrices both have full rank $k$ if we want to construct a $\UM{n,k}$ unit memory code. 

For a $\PUM{n,k}{k^{(1)}}$ partial unit memory code over $\F$, $\rank(\G^{(0)})=k$ and $\rank(\G^{(1)})=k^{(1)} < k$ has to hold. W.l.o.g., for PUM codes, we assume that the lowermost $k-k^{(1)}$ rows of $\G^{(1)} $ are zero and we denote: 
\begin{equation}\label{eq:pummatrix}
\G^{(0)} = \begin{pmatrix}
\mathbf G^{(00)}\\
 \mathbf G^{(01)}
\end{pmatrix}, 
 \quad \G^{(1)} = 
 \begin{pmatrix}
 \G^{(10)}\\
 \0
 \end{pmatrix},
\end{equation}
where $\G^{(00)}$ and $\G^{(10)}$ are $k^{(1)} \times n$ matrices {of full rank} and $\G^{(01)}$ is a {full-rank} $(k-k^{(1)})\times n$ matrix over $\F$. 
The encoding rule for each code block of a (P)UM code is hence given by 
\begin{equation}\label{eq:encoding_pum}
\c^{(i)} = \u^{(i)} \cdot \G^{(0)} + \u^{(i-1)} \cdot \G^{(1)}, \quad \forall i=1,2,\dots,
\end{equation}
where $\u^{(i)}$ and $\u^{(i-1)} \in \F^k$ for all $i$.
The memory of (P)UM codes is $\memory=1$.
The overall constraint length is $\constraintlength = k$ for UM codes and $\constraintlength = k^{(1)}$ for PUM codes.

\section{Distance Measures for Convolutional Codes in Rank Metric}\label{sec:conv_metrics}
In this section, we provide distance measures and upper bounds for convolutional codes based on a special rank metric, see also \cite{WachterSidBossZyb_PUMbasedGab}. This special rank metric---the \emph{sum rank metric}---was proposed by Nóbrega and Uch\^{o}a-Filho under the name "extended rank metric" in \cite{NobregaUchoa-2010Multishot} for multi-shot transmissions in a network. 

\subsection{Distance Parameters and Trellis Description}\label{subsec:distdef}
In \cite{NobregaUchoa-2010Multishot}, it is shown that the sum rank distance and the subspace distance of the modified lifting construction are related in the same way as the {rank distance} and the subspace distance of the {lifting construction}, see \cite{silva_rank_metric_approach} and also Lemma~\ref{lem:subspacecode_lifted_mrd}. Hence, the 
use of the sum rank metric for multi-shot network coding can be seen as the analog to using the rank metric for single-shot network coding.
 
The sum rank weight and distance are defined as follows.

\begin{definition}[Sum Rank Weight/Distance]\label{def:sum_Rank_Metric}
Let two vectors $\a, \b \in \F^{nN}$ be decomposed into $N$ subvectors of length $n$ such that: 
\begin{equation*}
\a = (\a^{(0)}\ \a^{(1)}\  \dots \ \a^{(N-1)}),\ \b = (\b^{(0)}\ \b^{(1)}\  \dots \ \b^{(N-1)}),
\end{equation*}
with $\a^{(i)},\b^{(i)}  \in \F^n$, $\forall i \in \intervallexcl{0}{N}$.
The sum rank weight of $\a$ is the sum of the ranks of the subvectors:
\begin{equation}\label{eq:defsumrankweight}
\wtsum(\a) \defeq 
\sum \limits_{i=0}^{N-1} \rk(\a^{(i)}).
\end{equation}
The sum rank distance between \vec{a} and \vec{b} is the sum rank weight of the difference of the vectors:
\begin{equation}\label{eq:defsumrankdist}
\distsum(\a,\b) 
\defeq \wtsum(\a-\b)
=\sum \limits_{i=0}^{N-1}\rk(\a^{(i)}-\b^{(i)}).
\end{equation}
\end{definition}
Since the rank distance is a metric (see e.g. \cite{Gabidulin_TheoryOfCodes_1985}), the sum rank distance is also a metric. 

An important measure for convolutional codes in Hamming metric is the free distance, and consequently, we define the \emph{free rank distance} in a similar way in the sum rank metric.
\begin{definition}[Free Rank Distance]\label{def:freedist}
The free rank distance of a convolutional code $\mycode{C}$ is the minimum sum rank distance \eqref{eq:defsumrankdist} between any two different codewords $\a,\b \in \mycode{C}$:
\begin{equation*}
\dfree
\defeq
\min_{\substack{\a,\b \in \mycode{C},\\ \a \neq \b}} \Big\lbrace \distsum(\a,\b)\Big \rbrace
=\min_{\substack{\a,\b \in \mycode{C},\\ \a \neq \b}} \left\lbrace \sum \limits_{i=0}^{N-1} \rk(\a^{(i)}-\b^{(i)})\right \rbrace.
\end{equation*}
\end{definition}
For a \emph{linear} convolutional code, the free rank distance is $\dfree =\min_{\substack{\a \in \mycode{C}, \a \neq \0}} \big\lbrace \wtsum(\a)\big \rbrace$.
Throughout this paper, we consider only linear convolutional codes.

Any convolutional code can be described by a minimal code trellis, which has a certain number of states and the input/output blocks are associated to the edges of the trellis.
The current state in the trellis of a (P)UM code over $\F$ can be associated with the vector $\s^{(i)} = \u^{(i-1)}\G^{(1)}$, see e.g., \cite{Justesen_BDDecodingUM}, and therefore there are $q^{mk^{(1)}}$ possible states.
We call the current state \emph{zero state} if $\s^{(i)} = \0$. 
A code sequence of a terminated (P)UM code with $N$ blocks can therefore be considered as a path in the trellis, which starts in the zero state and ends in the zero state after $N$ edges. 

The error-correcting capability of convolutional codes is determined by \emph{active distances}, a fact that will become obvious in view of our decoding algorithm in Section~\ref{sec:pum_decoding}. 
In the following, we define the \emph{active row/column/reverse column rank distances} analog to active distances in Hamming metric \cite{Thommesen_Justesen_BoundsUM,HoestJohannessonZigangirovZyablov-ActiveDistances_1999,Johannesson_Fund_Conv_Codes}. In the literature, there are different definitions of active distances in Hamming metric. Informally stated, for a $j$-th order active distance of $\mycode{C}$, we simply look at all sequences of length $j$, and require some conditions on the passed states in the minimal code trellis of $\mycode{C}$.

\begin{figure}[htb,scale = 1]
\centering
\scalebox{0.65}{\includegraphics{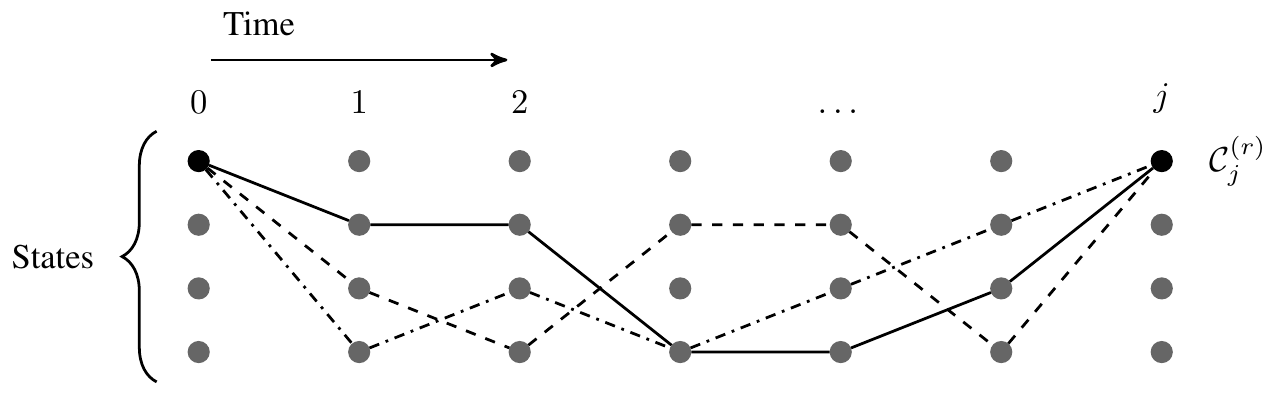}} 
\caption{The set $\ActRowDistSet{j}$: it consists of all codewords of $\mycode{C}$ having paths in the minimal code trellis which diverge from the zero state at depth $0$ and return to the zero state for the first time at depth $j$.}
\label{fig:setCrowdistance}
\end{figure}

Let $\ActRowDistSet{j}$ denote the set of all codewords in a convolutional code $\mycode{C}$, corresponding to paths in the minimal code trellis which diverge from the zero state at depth zero and return to the zero state for the \emph{first} time after $j$ branches at depth $j$. W.l.o.g., we assume that we start at depth zero, as we only consider time-invariant convolutional codes. This set is illustrated in Figure~\ref{fig:setCrowdistance}.

\begin{definition}[Active Row Rank Distance]\label{def:extrowrank}
The active row rank distance of order $j$ of a linear convolutional code is defined as
\begin{equation*}
\drow{j} \defeq \min_{\substack{\c \in \ActRowDistSet{j}}} \big\lbrace \wtsum(\c) \big\rbrace, \quad \forall j \geq 1.
\end{equation*}
\end{definition}
Clearly, for non-catastrophic encoders \cite{Johannesson_Fund_Conv_Codes}, the minimum of the active row rank distances of different orders is the same as the free rank distance, see Definition~\ref{def:freedist}:
$\dfree = \min_{\substack{j}}\left\lbrace \drow{j} \right\rbrace$.
The \emph{slope} of the active row rank distance is defined as follows.
\begin{definition}[Slope of Active Row Rank Distance]\label{def:slope}
The slope of the active row rank distance (Definition~\ref{def:extrowrank}) is
\begin{equation*}
\slope \defeq \lim_{\substack{j \rightarrow \infty}} \Bigg \lbrace \frac{\drow{j}}{j} \Bigg \rbrace.
\end{equation*}
\end{definition}
As in Hamming metric \cite[Theorem~1]{JordanPavZyb_MaxSlope_2004}, \cite[Theorem~2.7]{Jordan_Diss_WovenConv}, the active row rank distance of order $j$ can be lower bounded by a linear function $\drow{j} \geq \max \lbrace j \cdot \slope +\beta, \dfree \rbrace$ for some $\beta \leq \dfree$.

\begin{figure}[htb]
\centering
\scalebox{0.85}{\includegraphics{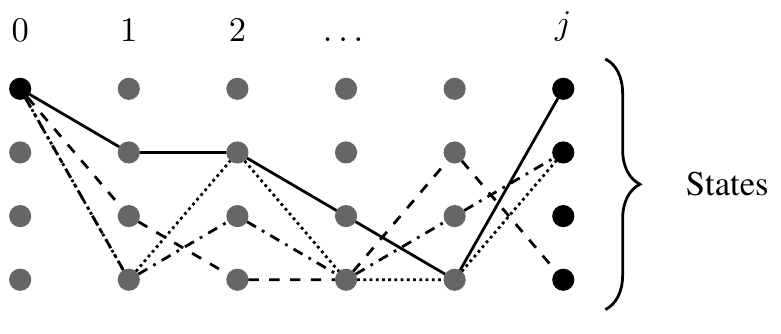}} 
\caption{The set $\ActColDistSet{j}$: all codewords of $\mycode{C}$ diverging from the zero state at depth $0$, where no zero states between depths $0$ and $j$ are allowed. \label{fig:extendedcoldist}}
\end{figure}

\begin{figure}[htb]
\centering
\scalebox{0.85}{\includegraphics{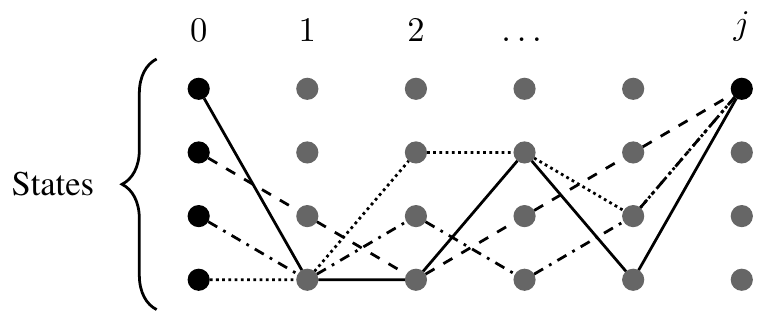}} 
\caption{The set $\ActRevColDistSet{j}$: all codewords of $\mycode{C}$ ending in the zero state at depth $j$, where no zero states between depths $0$ and $j$ are allowed. \label{fig:extendedcoldist_rev}}
\end{figure}

Similar to Hamming metric, we can introduce an active column rank distance and an active reverse column rank distance.
Let $\ActColDistSet{j}$ denote the set of all words in the trellis of length $j$ blocks, leaving the zero state at depth zero and ending in any state at depth $j$ and let 
$\ActRevColDistSet{j}$ denote the set of all words starting in any state at depth zero and ending in the zero state in depth $j$, 
both without zero states in between (see Figures~\ref{fig:extendedcoldist} and \ref{fig:extendedcoldist_rev}).
The active column rank distance and the active reverse column rank distance are then defined by:
\begin{align}
\dcol{j} &\defeq \min_{\substack{\c \in \ActColDistSet{j}}} \big\lbrace\wtsum (\c) \big\rbrace,\quad \forall j \geq 1. \label{eq:def_actcol_dist}\\
\drevcol{j} &\defeq \min_{\substack{\c \in \ActRevColDistSet{j}}} \big\lbrace \wtsum(\c)\big\rbrace,\quad \forall j \geq 1.\label{eq:def_revcol_dist}
\end{align}


\subsection{Upper Bounds on Distances of (P)UM Codes}\label{subsec:upperbound_pum}
In the following, we recall upper bounds on the free rank distance $\dfree$ (Definition~\ref{def:freedist}) and the slope $\slope$ (Definition~\ref{def:slope}) for UM and PUM codes based on the sum rank metric \eqref{eq:defsumrankweight}, \eqref{eq:defsumrankdist}. 
The derivation of the bounds uses known bounds for (P)UM codes in Hamming metric \cite{Lee_UnitMemory_1976,Lauer_PUM_1979,Pollara_FiniteStateCodes}.

\begin{corollary}[Upper Bounds {\cite[Corollary~1]{WachterSidBossZyb_PUMbasedGab}}]\label{cor:upperbounds}
For a $\UM{n,k}$ code, where $\constraintlength=k$, the free rank distance is bounded by:
\begin{equation}\label{eq:umupperbound}
\dfree \leq 2n-k+1.
\end{equation}
For a $\PUM{n,k}{k^{(1)}}$ code, where $\constraintlength=k^{(1)} < k$, the free rank distance is bounded by:
\begin{equation}\label{eq:pumupperbound}
\dfree \leq n-k+\constraintlength+1.
\end{equation}
For both, UM and PUM codes, the slope is bounded by:
\begin{equation}\label{eq:slopeupperbound}
\slope \leq n-k.
\end{equation}
\end{corollary}

\section{Construction of Convolutional Codes in Rank Metric}\label{sec:pum_constructions}
This section provides a construction of (P)UM codes whose submatrices of the generator matrix define Gabidulin codes. 
In the first step (Section~\ref{subsec:low-rate-constr}), we adapt the construction from \cite{DettmarSorger_BMDofUM} in Hamming metric to rank metric, yielding low-rate (P)UM codes. 
Later in Section~\ref{subsec:arb-rate-constr}, as in \cite{WachterStinner_BMDArbitraryRates_2012_conf}, we extend the construction to arbitrary code rates.

\subsection{Low-Rate Code Construction}\label{subsec:low-rate-constr}
The following definition provides our code construction.
\begin{definition}[(P)UM Code based on Gabidulin Code]\label{def:pumgab_genmat}
Let $k + k^{(1)} \leq n \leq m$, where $k^{(1)} \leq k$. Further, let $g_0,g_1,\dots, g_{n-1} \in \F$ be linearly independent over $\Fq$.


For $k^{(1)} \leq k$, we define a $\PUM{n,k}{k^{(1)}}$ code, respectively a $\UM{n,k}$ code, over $\F$ by
a zero-forced terminated generator matrix $\G$ as in \eqref{eq:genmat_termin} with $\memory=1$.
We use the $k \times n $ submatrices $\G^{(0)}$ and $\G^{(1)}$:
\begin{equation}\label{eq:constrG0}
\G^{(0)}=
\begin{pmatrix}\mathbf G^{(00)}\\\hline\\[-2ex] \mathbf G^{(01)}\end{pmatrix}
=\begin{pmatrix}
g_{0} & g_{1} & \dots& g_{n-1}\\
g_{0}^{[1]} & g_{1}^{[1]} & \dots& g_{n-1}^{[1]}\\
\vdots&\vdots&\ddots&\vdots\\
g_{0}^{[k^{(1)}-1]} & g_{1}^{[k^{(1)}-1]} & \dots& g_{n-1}^{[k^{(1)}-1]}\\[0.5ex]
\hline\\[-2ex]
g_{0}^{[k^{(1)}]} & g_{1}^{[k^{(1)}]} & \dots& g_{n-1}^{[k^{(1)}]}\\
g_{0}^{[k^{(1)}+1]} & g_{1}^{[k^{(1)}+1]} & \dots& g_{n-1}^{[k^{(1)}+1]}\\
\vdots&\vdots&\ddots&\vdots\\
g_{0}^{[k-1]} & g_{1}^{[k-1]} & \dots& g_{n-1}^{[k-1]}\\
\end{pmatrix},
\end{equation}
and 
\begin{align}
\G^{(1)}&=
\begin{pmatrix}\mathbf G^{(10)}\\\hline\\[-2ex] \0\end{pmatrix}\label{eq:constrG1}\\
&=\begin{pmatrix}
g_{0}^{[k]} & g_{1}^{[k]} & \dots& g_{n-1}^{[k]}\\
g_{0}^{[k+1]} & g_{1}^{[k+1]} & \dots& g_{n-1}^{[k+1]}\\
\vdots&\vdots&\ddots&\vdots\\
g_{0}^{[k+k^{(1)}-1]} & g_{1}^{[k+k^{(1)}-1]} & \dots& g_{n-1}^{[k+k^{(1)}-1]}\\[0.5ex]
\hline\\[-2ex]
&  &  & \\
&&\hspace{-1.9cm}\smash{\clap{\resizebox{0.35cm}{!}{$0$}}}&\nonumber\\
\end{pmatrix}.
\end{align}
\end{definition}

Table~\ref{tab:subcodes_pum} denotes some Gabidulin codes, which are defined by submatrices of $\G$, their minimum rank distances and their block rank-metric error-erasure BMD decoders (realized e.g., by the decoders from \cite{GabidulinPilipchuck_ErrorErasureRankCodes_2008,silva_rank_metric_approach}).
These BMD decoders decode correctly if \eqref{eq:errorerasurecond} is fulfilled for the corresponding minimum rank distance.
If we consider unit memory codes with $k = k^{(1)}$, then $d_{00} = d_{10} = n-k+1$, $\da = n-2k+1$ and $d_{01} = \infty$, since $\G^{(01)}$ does not exist.

\begin{table*}[ht]
    \caption{Submatrices of (P)UM code from Definition~\ref{def:pumgab_genmat} and their block codes.}
    \label{tab:subcodes_pum}
    \centering
    \renewcommand{\arraystretch}{1.4}
	\begin{tabular}{p{0.13\textwidth} p{0.08\textwidth} p{0.14\textwidth} p{0.18\textwidth} p{0.08\textwidth}}
		\toprule	
		Generator\newline matrix	& Code\newline notation & Code\newline parameters & Minimum rank\newline distance & BMD\newline decoder	\\
		\midrule
		$\G^{(0)}$ & $\mycode{C}_0$ &$\Gab{n,k}$& $d_0 = n-k+1$ & \BMD{\mycode{C}_0}\\[0.5ex]
		$\left(\begin{smallmatrix}\mathbf G^{(01)}\\\mathbf G^{(10)}\end{smallmatrix}\right)$ & $\mycode{C}_1$ &$\Gab{n,k}$& $d_1 = n-k+1$ & \BMD{\mycode{C}_1}\\[0.5ex]
		$\G^{(01)}$ & $\mycode{C}_{01}$ &$\Gab{n,k-k^{(1)}}$& $d_{01} = n-k+k^{(1)}+1$ & \BMD{\mycode{C}_{01}}\\[0.5ex]
		$\mathbf G_{\sigma} = \left(\begin{smallmatrix}\mathbf G^{(00)}\\\mathbf G^{(01)}\\\mathbf G^{(10)}\end{smallmatrix}\right)$ & $\Ca$ &$\Gab{n,k+k^{(1)}}$& $\da = n-k-k^{(1)}+1$ & $\BMD{\Ca}$\\[0.5ex]
		\bottomrule
    \end{tabular}
\end{table*}

{
To show that the generator matrix $\G$ of Definition~\ref{def:pumgab_genmat} 
 is in minimal basic encoding form, see \cite[Definitions~4 and 5]{Forney_ConvolutionalCodesAlg} and \cite{JohannessonWan-ALinearAlgebraicApproachToMinimalConvolutionalEncoders}, let us slightly generalize Theorem 6 from \cite{JohannessonWan-ALinearAlgebraicApproachToMinimalConvolutionalEncoders} for the case of arbitrary finite field $\Fq$ as follows.  Let  $[\G(D)]_h$ be a matrix over $\Fq$ having the leading coefficient of $g_{ij}(D)$ in position $(i,j)$ if $\deg g_{ij}(D)= \nu_i$ and $0$ otherwise.
\begin{lemma}\label{lemma2}
A basic encoding matrix $\G(D)$ over $\F_q[D]$ is minimal basic if and only if  $[\G(D)]_h$ has full rank.
\end{lemma}
\begin{IEEEproof}
According to \cite[Definition 5]{Forney_ConvolutionalCodesAlg}, a basic convolutional generator $k\times n$ matrix $\G(D)$ is minimal iff its overall constraint length $\nu$ is equal to the maximum degree  $\eta$  of its $k\times k$ subdeterminants, $\eta=\nu$. Select a $k\times k$ submatrix $\G'(D)$ of $\G(D)$ with  $\deg \det \G'(D)=\eta$. From the Leibniz formula for calculating the determinant  $p(D) = \det \G'(D)$ it follows that $\eta \le \nu$ and the coefficient $p_\nu$ in $p(D)$ is $p_\nu=\det [\G'(D)]_h$. Hence, there exists submatrix $\G'(D)$ of $\G(D)$ with $\deg \det \G'(D) = \nu$ iff $p_\nu=\det [\G'(D)]_h \ne 0$, which is iff the matrix $[\G(D)]_h$ has full rank, and the statement of the lemma follows. 
\end{IEEEproof}}

\begin{theorem}[Minimal Basic Encoding Form]\label{lem:pum_constr_G_minBasicEncForm}
 Let a (P)UM code based on Gabidulin codes be defined by its generator matrix $\G$ as in Definition~\ref{def:pumgab_genmat}. 
 Then, $\G$ is in minimal basic encoding form, see \cite[Definitions~4 and 5]{Forney_ConvolutionalCodesAlg}. 
\end{theorem}
\begin{IEEEproof}
First, $\G$ is in \emph{encoding} form since $\G^{(0)}$ is a $q$-Vandermonde matrix and therefore has full rank \cite{Lidl-Niederreiter:FF1996}.

Second, we show that $\G$ is in \emph{basic} form. 
According to \cite[Definition 4]{Forney_ConvolutionalCodesAlg}, $\mathbf G(D)$ is basic if it is polynomial and if there exists a polynomial right inverse $\mathbf G^{-1}(D)$, such that $\mathbf G(D) \cdot \mathbf G^{-1}(D)=\mathbf I_{k\times k}$. 
By definition, $\mathbf G(D)$ is polynomial. A polynomial right inverse exists if and only if $\mathbf G(D)$ is non-catastrophic and hence if the slope is $\slope >0$ \cite[Theorem~A.4]{Dettmar_PUM_PhD_1994}. The slope is calculated later in Corollary~\ref{cor:dfreeslope_G}, proving that $\slope>0$.

{
Third, we show that $\mathbf G(D)$ is \emph{minimal}.
Indeed, the matrix
\begin{equation*}
[\G(D)]_h=\left(
\begin{array}{c}
\mathbf G^{(10)}\\
\mathbf G^{(01)}\\
\end{array}
\right)
\end{equation*} 
has full rank by Definition \ref{def:pumgab_genmat} and minimality follows from Lemma~\ref{lemma2}.
 } 
\end{IEEEproof}

In the following, we calculate the active row, column and reverse column rank distances (Definition~\ref{def:extrowrank} and Equations~\eqref{eq:def_actcol_dist}, \eqref{eq:def_revcol_dist}) by cutting the generator matrix of the PUM code from Definition~\ref{def:pumgab_genmat} into parts. 
Pay attention that each code block of length $n$ can be seen as a codeword of $\Ca$. 

\begin{theorem}[Lower Bound on Active Distances]\label{the:LBAD}
Let $k + k^{(1)} \leq n \leq m$, where $k^{(1)} \leq k$.
Let $\mycode{C}$ be a $\UM{n,k}$, respectively $\PUM{n,k}{k^{(1)}}$, code over $\F$ as in Definition~\ref{def:pumgab_genmat}.

Then,
\begin{align*}
 \drow{1} \geq\drdes{1}  &=  d_{01},\\ 
 \drow{j} \geq\drdes{j} &= d_0 + (j-2) \cdot \da + d_1, \quad \forall j\geq 2,\\
\dcol{j} \geq \dcdes{j} &= d_0 + (j-1) \cdot \da, \quad \forall  j \geq 1,\\
\drevcol{j} \geq \drcdes{j} &=  (j-1) \cdot \da + d_1,\quad \forall j \geq 1,
\end{align*}
where $d_{01}=n-k + k^{(1)}+1$ for $k^{(1)} < k$ and $d_{01}=\infty$ for $k^{(1)}=k$, $d_0 = d_1 = n-k+1$, $\da = n-k-k^{(1)}+1$.
\end{theorem}
\begin{IEEEproof}
For the estimation of the active row rank distance,
the encoder starts in the zero state  hence, $\u^{(-1)}=\0$.
For the first order active row distance $\drow{1}$, we look at all code sequences of the form $(\dots \ \0 \ \c^{(0)} \ \0 \ \dots)$, which is only possible if $\u^{(0)}=(0 \ \dots \ 0 \ u^{(0)}_{k^{(1)}} \ \dots \ u^{(0)}_{k-1})$ and $\u^{(i)} = \0$, $\forall i \geq 1$. 
In this case, $\c^{(0)} \in \mycode{C}_{01}$ with distance $d_{01}$, and the encoder returns immediately to the zero state. For the UM case, $\u^{(0)} = \0$ and the only codeword in $\ActRowDistSet{0}$ is the all-zero codeword and thus, $\drow{1} = \infty$.

For higher orders of $\drow{j}$, we have to consider all code sequences, starting with 
$\c^{(0)} \in \mycode{C}_0$ (since $\u^{(-1)}=\0$), followed by $j-2$ non-zero codewords of $\Ca$ and 
one final code block, resulting from $\u^{(j-1)}=(0 \ \dots \ 0 \  u^{(j-1)}_{k^{(1)}} \ \dots \ u^{(j-1)}_{k-1})$ and for the UM case $\u^{(j-1)}=\0$. 
For the UM and the PUM case, the block $\u^{(j-2)}$ is arbitrary, therefore 
 $\c^{(j-1)} = \u^{(j-1)} \cdot \G^{(0)} + \u^{(j-2)} \cdot \G^{(1)} \in \mycode{C}_1$. 
 
For the estimation of $\dcol{j}$, the encoder starts in the zero state but ends in any state. 
Thus, $\c^{(0)} \in \mycode{C}_0$ is followed 
 by $j-1$ arbitrary information blocks resulting in codewords from $\Ca$.\\ 
For the active reverse column rank distances, we start in any block, hence, all first $j-1$ blocks
 are from $\Ca$. The last block is from $\mycode{C}_1$ in order to end in the zero state. 
\end{IEEEproof}

We call $\drdes{j}$, $\dcdes{j}$, $\drcdes{j}$ \emph{designed} active distances in the following since they are lower bounds on $\drow{j}$, $\dcol{j}$, $\drevcol{j}$.

\begin{corollary}[Free Rank Distance and Slope]\label{cor:dfreeslope_G}
Let $k + k^{(1)} \leq n \leq m$, where $k^{(1)} \leq k$.
Let $\mycode{C}$ be a $\UM{n,k}$, respectively $\PUM{n,k}{k^{(1)}}$, code over $\F$ as in Definition~\ref{def:pumgab_genmat}.

Then, its free rank distance $\dfree$ for $k^{(1)} = k$ is
\begin{equation*}
\dfree  \geq \min_{\substack{j}} \left\lbrace \drdes{j}\right\rbrace = d_0+d_1 = 2(n-k+1),\\
\end{equation*}
and for $k^{(1)} < k$:
\begin{equation*}
\dfree  = \min_{\substack{j}} \left\lbrace \drdes{j}\right\rbrace = d_{01} = n-k + k^{(1)}+1 = n-k + \constraintlength+1.\\
\end{equation*}
The slope $\slope$ of $\mycode{C}$ for both cases is:
\begin{equation*}
\slope \geq \lim_{\substack{j \rightarrow \infty}} \Bigg \lbrace \frac{\drdes{j}}{j} \Bigg \rbrace = \da =  n-k-k^{(1)}+1.
\end{equation*}
\end{corollary}
Thus, for any $k^{(1)} < k$, the construction attains the upper bound on the free rank distance of PUM codes \eqref{eq:pumupperbound}.
When $k^{(1)} = k = 1$, we attain the upper bound on the free rank distance of UM codes, see \eqref{eq:umupperbound}.
For $k^{(1)}=1 \leq k$, the upper bound on the slope is attained.

If we compare this to the construction from \cite{WachterSidBossZyb_PUMbasedGab},
we see that both constructions attain the upper bound on the free rank distance for $k^{(1)} < k$. It depends on the explicit values of $n$, $k$ and $k^{(1)}$, which construction has a higher slope.

The construction based on the parity-matrix from \cite{WachterSidBossZyb_PUMbasedGab} requires that $R=k/n \geq \memoryH/(\memoryH+1)$, where $\memoryH \geq 1$, and provides therefore a high-rate code construction, whereas the construction based on the generator matrix (Definition~\ref{def:pumgab_genmat}) results in a low-rate code since $k + k^{(1)} \leq n$ has to hold.

\subsection{Construction of Arbitrary Code Rate}\label{subsec:arb-rate-constr}
In the sequel, we outline how to extend the construction from Definition~\ref{def:pumgab_genmat} to arbitrary code rates. Compared to the high-rate construction from \cite{WachterSidBossZyb_PUMbasedGab}, the advantage is that we are able to decode this code construction efficiently (see Section~\ref{sec:pum_decoding}).
We apply the same strategy to extend the construction from Definition~\ref{def:pumgab_genmat} to arbitrary code rates as in \cite{WachterStinner_BMDArbitraryRates_2012_conf} in Hamming metric. Further, we use the same notations for the matrices as in the previous section, but with an additional prime symbol for each matrix (e.g., $\G$ becomes $\G^\prime$).


So far, we have defined the code $\Ca$ as a $\Gab{n,k+k^{(1)}}$ code with $\da=n-k-k^{(1)}+1$, where $k+k^{(1)} \leq n$ (see also Table~\ref{tab:subcodes_pum}).
Overcoming the restriction $k+k^{(1)} \leq n$ would enable us to choose an arbitrary code rate $R = k/n$ of the convolutional code \mycode{C} for any fixed $k^{(1)}$. However, at the same time, if $k+k^{(1)}>n$, there have to be linearly dependent rows in $\left(\begin{smallmatrix}
\G^{(0)\prime}\\
\G^{(1)\prime}
\end{smallmatrix}\right)$.

Therefore, we define these matrices such that $\varphi$ denotes the number of rows which are contained (amongst others) in both, $\G^{(0)\prime}$ and $\G^{(1)\prime}$.
We define a full-rank $(k+k^{(1)}-\varphi)\times n$ matrix $\mathbf G_{all}'$ by 
\begin{equation}\label{eq:Gtot}
\mathbf G_{all}^\prime =
\begin{pmatrix}
 \mathbf A\\
 \mathbf \Phi\\
 \mathbf \G^{(01)\prime}\\
 \mathbf B
\end{pmatrix},
\end{equation}
where $\mathbf A$ and $\mathbf B$ are in $\F^{(k^{(1)}-\varphi)\times n}$, $\mathbf \Phi$ is in $\F^{\varphi\times n}$ and $\mathbf G^{(01)\prime} \in \F^{(k-k^{(1)})\times n}$ such that $\mathbf G_{all}^\prime$ is a generator matrix of a $\Gab{n,k+k^{(1)}-\varphi}$ code of minimum rank distance $d_{all} = \da^\prime = n-k-k^{(1)}+\varphi + 1$. Clearly, $k+k^{(1)}-\varphi\leq n$ and $\da^\prime \leq n$ have to hold.
Since $\mathbf G_{all}'$ defines a Gabidulin code, any submatrix of consecutive rows defines a Gabidulin code as well.
Based on the definition of $\mathbf G_{all}^\prime$, our generalized PUM code construction is given as follows.

\begin{definition}[Generalized (P)UM Code Construction]\label{def:pumarbitrary}
Let $k + k^{(1)} -\varphi \leq n \leq m$, where $\varphi < k^{(1)} \leq k$.
Further, let $\mathbf G_{all}^{\prime}$ be as in \eqref{eq:Gtot}, defining an $\Gab{n,k+k^{(1)}-\varphi}$ code.
Our rate $k/n$ (P)UM code is defined by the following submatrices:
\begin{equation}\label{eq:G12}
\G^{(0)\prime} =\!
\begin{pmatrix}
 \mathbf G^{(00)\prime}\\
 \mathbf G^{(01)\prime}
\end{pmatrix}\!=\!\begin{pmatrix}
 \mathbf A\\
 \mathbf \Phi\\
 \mathbf \G^{(01)\prime}\\
\end{pmatrix}, 
\ \G^{(1)\prime} = \!
\begin{pmatrix}
 \G^{(10)\prime}\\
 \0
\end{pmatrix}\!=\!
\begin{pmatrix}
 \mathbf \Phi\\
 \mathbf B\\
 \0
\end{pmatrix}\!,
\end{equation}
\end{definition}
{where $\0$ is the all-zero matrix.}
We restrict $\varphi<k^{(1)}$ since otherwise all rows of $\G^{(1)\prime}$ are rows of $\G^{(0)\prime}$. Further, any code rate $k/n$ in combination with any $k^{(1)}$ is feasible with this restriction since $k+1 \leq k+ k^{(1)}-\varphi \leq n$ and hence, we have only the trivial restriction $k < n$.

{\begin{theorem}
The generator matrix $\G^{\prime}(D)=\G^{(0)\prime}+\G^{(1)\prime}D$ from Definition~\ref{def:pumarbitrary} of the Generalized (P)UM code is in minimal basic encoding form. 
\end{theorem}
\begin{IEEEproof}
The proof is similar to the one of Theorem~\ref{lem:pum_constr_G_minBasicEncForm}. The matrix $\G^{\prime}$ is minimal as the matrix 
\begin{equation*}
[\G^{\prime}(D)]_h=\left(
\begin{array}{c}
\mathbf \Phi\\
\mathbf B\\
\mathbf G^{(01)\prime}\\
\end{array}
\right)
\end{equation*} 
has full rank by Definition \ref{def:pumarbitrary}, since it is a submatrix of generator matrix of a Gabidulin code, and minimality follows from Lemma~\ref{lemma2}.
\end{IEEEproof}}

To calculate the active distances of the generalized code construction from Definition~\ref{def:pumarbitrary}, we need to take into account that consecutive \emph{non-zero} information blocks can result in \emph{zero} code blocks due to the linear dependencies in the rows of $\G^{(0)\prime}$ and $\G^{(1)\prime}$.
This is shown in the following example.

\begin{example}[Zero Code Block]
Let two consecutive information blocks $\u^{(j-1)}$, $\u^{(j)} \in \F^k$ be:
\begin{align*}
\u^{(j-1)}&=(u^{(j-1)}_{0} \ \dots \ u^{(j-1)}_{\varphi-1} \ 0 \ \dots \ 0),\\
\u^{(j)}&=(0 \ \dots \ 0 \ u^{(j)}_{k^{(1)}-\varphi} \ \dots \ u^{(j)}_{k^{(1)}-1} \ 0 \ \dots \ 0).
\end{align*}
By encoding $\c^{(j)}$, we obtain
\begin{align*}
\c^{(j)} 
&= \u^{(j)}\cdot \begin{pmatrix}
 \mathbf A\\
 \mathbf \Phi\\
 \mathbf \G^{(01)\prime}
\end{pmatrix}
+\u^{(j-1)} \cdot\begin{pmatrix}
 \mathbf \Phi\\
 \mathbf B\\
 \0
\end{pmatrix}\\
&=\left( (u^{(j)}_{k^{(1)}-\varphi} \ \dots \ u^{(j)}_{k^{(1)}-1}) 
+ (u^{(j-1)}_{0} \ \dots \ u^{(j-1)}_{\varphi-1}) \right) \cdot \mathbf\Phi.
\end{align*}
If $(u^{(j)}_{k^{(1)}-\varphi} \ \dots \ u^{(j)}_{k^{(1)}-1})=- (u^{(j-1)}_{0} \ \dots \ u^{(j-1)}_{\varphi-1})$,
we obtain an all-zero code block $\c^{(j)}=\0$ although $\u^{(j-1)}, \u^{(j)} \neq \0$.
\end{example}

However, in the same way as in \cite[Lemma~1]{WachterStinner_BMDArbitraryRates_2012_conf}, it can be shown that the maximum number of consecutive zero blocks is bounded from above by $\ell = \lceil\varphi/(k^{(1)}-\varphi) \rceil$. 
Hence, after at most $\ell$ zero code blocks, there has to be (at least) one non-zero code block and the slope can be lower bounded by
\begin{align}
\slope^\prime\geq\frac{\da^\prime}{\ell+1}=\frac{n-k-k^{(1)}+\varphi+1}{\lceil\tfrac{k^{(1)}}{k^{(1)}-\varphi}\rceil}.
\end{align}
This provides the following extended distances:
\begin{align}
 \drowprime{1} \geq\drdesprime{1}  &=  d_{01},\label{eq:ext_dist_arb_rate}\\ 
 \drow{j} \geq\drdesprime{j} &= d_0 + (j-2) \cdot \slope^\prime + d_1, \quad \forall j\geq 2,\nonumber\\
\dcolprime{j} \geq \dcdesprime{j} &= d_0 + (j-1) \cdot \slope^\prime, \quad \forall  j \geq 1,\nonumber\\
\drevcolprime{j} \geq \drcdesprime{j} &=  (j-1) \cdot \slope^\prime + d_1,\quad \forall j \geq 1,\nonumber
\end{align}
which reduces to the distances of Theorem \ref{the:LBAD} for $\ell=0$.
Note that $d_0$, $d_{01}$ and $d_1$ are independent of $\varphi$ and therefore the same as in Table~\ref{tab:subcodes_pum}.
We see that there is a trade-off between the code rate and the extended distances; namely, the higher the code rate, the higher $\varphi$ (for fixed $k^{(1)}$), and the lower $\slope^\prime$ and the lower the extended distances (for constant $\varphi-k$).

\section{Error-Erasure Decoding of PUM Gabidulin Codes}\label{sec:pum_decoding}
This section provides an efficient error-erasure decoding algorithm for our construcion of (P)UM codes as in Definition~\ref{def:pumgab_genmat},
using the block rank-metric decoders of the underlying Gabidulin codes in Table~\ref{tab:subcodes_pum}. We explain the general idea, prove its correctness and show how to generalize the decoding algorithm to the arbitrary-rate construction from Definition~\ref{def:pumarbitrary}.

\subsection{Bounded Row Distance Condition and Decoding Algorithm}
We consider the terminated generator matrix of a (P)UM code as in \eqref{eq:genmat_termin} and therefore, each codeword has length $N+\memory = N+1$.
Let the received sequence $\r = \c + \e= (\r^{(0)} \ \r^{(1)} \ \dots \ \r^{(N)}) \in \F^{n(N+1)}$ be given and let the matrix sequence $\R = (\R^{(0)} \ \R^{(1)} \ \dots \ \R^{(N)}) \in \Fq^{m \times n(N+1)}$ denote the matrix representation of $\r$, where $\R^{(i)} = \extsmallfieldinput{\r^{(i)}}$, $\forall i \in \intervallincl{0}{N}$. 

Let $\R^{(i)}=\C^{(i)}+\E^{(i)}$, for all $i \in \intervallincl{0}{N}$, where $\R^{(i)} \in \Fq^{m \times n}$ can be decomposed as in \eqref{eq:decomp_errrorerasures}, including $t^{(i)}$ errors, $\numbRowErasures^{(i)}$ row erasures and $\numbColErasures^{(i)}$ column erasures in rank metric. 

Analog to Justesen's definition in Hamming metric \cite{Justesen_BDDecodingUM}, we define a 
bounded (row rank) distance decoder for convolutional codes in rank metric, incorporating additionally erasures.
\begin{definition}[BRD Error--Erasure Decoder]\label{def:bmddecoder}
Given a received sequence $\r = \c+\e \in \F^{n(N+1)}$, 
a bounded row distance (BRD) error-erasure decoder in rank metric for a convolutional code $\mycode{C}$ guarantees 
to find the code sequence $\c \in \mycode{C}$ if 
\begin{align}
&\sum\limits_{h=i}^{i+j-1}\Big(2 \cdot t^{(h)} + \numbRowErasures^{(h)} + \numbColErasures^{(h)}\Big) <\drdes{j}, \label{eq:bmdcond}\\
&\qquad \forall i \in \intervallincl{0}{N}, j \in \intervallincl{0}{N-i+1},\nonumber
\end{align}
where $t^{(h)}$, $\numbRowErasures^{(h)}$, $\numbColErasures^{(h)}$ denote the number of errors, row and column erasures in block $\E^{(h)}=\extsmallfieldinput{\e^{(h)}} \in\F^n$ as in \eqref{eq:decomp_errrorerasures}.
\end{definition}
In Algorithm~\ref{algo:pum}, we present such a BRD rank-metric error-erasure decoder for (P)UM codes constructed as in Definition~\ref{def:pumgab_genmat}. 
It is a generalization of the Dettmar--Sorger algorithm \cite{DettmarSorger_BMDofUM} 
to rank metric and to error-erasure correction. 
The generalization to error-erasure decoding can be done in a similar way in Hamming metric.


The main idea of Algorithm~\ref{algo:pum} is to take advantage of the algebraic structure of the underlying block codes 
and their efficient decoders (see Table~\ref{tab:subcodes_pum}). We use the outputs of these block decoders to build a reduced trellis, which has only very few states at every depth. 
As a final step of our decoder, the well-known Viterbi algorithm is applied to this reduced trellis. Since there are only a few states in the trellis, the Viterbi algorithm has quite low complexity.

The first step of Algorithm~\ref{algo:pum} is to decode $\r^{(i)}$, $\forall i\in \intervallexcl{1}{N}$, with $\BMD{\Ca}$, since each code block $\c^{(i)}$ is a codeword of $\Ca$, $\forall i\in \intervallexcl{1}{N}$. This decoding is guaranteed to be successful if $2t^{(i)}+\numbRowErasures^{(i)}+\numbColErasures^{(i)} < \da$.
Because of the termination, the first and the last block can be decoded in the codes $\mycode{C}_0$ and $\mycode{C}_{01}$, respectively, which have a higher minimum rank distance than $\Ca$.
Let $\c^{(i)\prime}$, for all $i\in \intervallincl{0}{N}$, denote the result of this decoding when it is successful.

\printalgoIEEEWidth{ \vspace{1ex}
 \caption{\newline$\c \leftarrow$ \textsc{BoundedRowDistanceDecoderPUM}$\big(\r\big)$}
 \label{algo:pum}
 \SetKwInput{KwIn}{\underline{Input}}
 \SetKwInput{KwOut}{\underline{Output}}
 \SetKwInput{KwIni}{\underline{Initialize}}
\DontPrintSemicolon
\SetAlgoVlined
\LinesNumbered
\KwIn{Received sequence $\mathbf{r} = (\r^{(0)} \ \r^{(1)} \ \dots \ \r^{(N)}) \in \Fq^{n(N+1)}$}
\BlankLine
\textbf{Step 1: }Decode $\r^{(0)}$ with \BMD{\mycode{C}_0}\\
\hspace{7.7ex}Decode $\mathbf{r}^{(i)}$ with \BMD{\Ca}, for all $i \in \intervallexcl{1}{N}$\;
\hspace{7.7ex}Decode $\r^{(N)}$ with \BMD{\mycode{C}_{01}}\\
\hspace{7.7ex}Assign metric $m^{(i)}$ as in \eqref{eq:defmetric}, for all $i \in \intervallincl{0}{N}$\;
\BlankLine
\textbf{Step 2: }For all found $\c^{(i)}$: decode $\pumforward{i}$ steps forward with\\ \hspace{25.3ex}\BMD{\mycode{C}_0},\;
\hspace{24.5ex}  decode $\pumbackward{i}$ steps backward with\\ \hspace{24.5ex} \BMD{\mycode{C}_1}\;
\BlankLine
\textbf{Step 3: }For all found $\c^{(i)}$:  decode $\r^{(i+1)}$ with \BMD{\mycode{C}_{01}}\;
\hspace{7.5ex}Assign metric $m^{(i)}$ as in \eqref{eq:defmetric_step3}, for all $i \in \intervallincl{0}{N}$\;
\BlankLine
\textbf{Step 4: }Find complete path with smallest sum rank metric\\ \hspace{7.5ex} using the Viterbi algorithm\;
\BlankLine
\KwOut{Codeword sequence $\c= (\c^{(0)} \ \c^{(1)} \ \dots \ \c^{(N)}) \in \F^{n(N+1)}$}
 \vspace{1ex}}{0.52}

For all $i\in \intervallincl{0}{N}$, we draw an edge in a reduced trellis with the following edge metric: \\[-4ex]

\begin{small}
\begin{align}
\hspace{-18ex}m^{(i)}\!=\!\!\begin{cases}
                  \, \rk(\r^{(i)}-\c^{(i)\prime}), &\parbox[t]{0.12\textwidth}{
                  \text{if \BMD{\mycode{C}_0} ($i=0$), \BMD{\Ca}} \text{($i \in \intervallexcl{1}{N}$),   \BMD{\mycode{C}_{01}} ($i=N$)} successful} 
                  \\[5ex]
		  \bigg\lfloor\frac{\da+1+\numbRowErasures^{(i)}+\numbColErasures^{(i)}}{2}\bigg\rfloor, \quad&\text{else.}
                 \end{cases}\nonumber\\[-4ex]
                 \label{eq:defmetric}
\end{align}
\end{small}

Notice that the metric for the successful case is always smaller than the metric for the non-successful case since
\begin{equation*}
\rk(\r^{(i)}-\c^{(i)\prime}) =t^{(i)}+\numbRowErasures^{(i)}+\numbColErasures^{(i)} 
\leq \bigg\lfloor \frac{\da+1+\numbRowErasures^{(i)}+\numbColErasures^{(i)}}{2}\bigg\rfloor -1.
\end{equation*}

If the block error-erasure decoder $\BMD{\Ca}$ decodes correctly, the result is $\c^{(i)\prime} = \u^{(i)}\G^{(0)} + (u_0^{(i-1)} \ u_1^{(i-1)} \ \dots \ u_{k^{(1)}-1}^{(i-1)})\cdot \G^{(10)}$. 
Since the minimum distance is
$\da \geq 1$,
we can reconstruct the whole information vector
$\u^{(i)} = (u_0^{(i)} \ u_1^{(i)} \dots \ u_{k-1}^{(i)})$ as well as the part of the previous information vector, i.e., $(u_0^{(i-1)} \ u_1^{(i-1)} \dots \ u_{k^{(1)}-1}^{(i-1)})$.

Assume, we reconstructed $\u^{(i)} $ and $(u_0^{(i-1)} \ u_1^{(i-1)} \ \dots \ u_{k^{(1)}-1}^{(i-1)})$ in Step~1, then 
we can calculate:
\begin{align}
 &\r^{(i+1)}\! - (u_0^{(i)}\ u_1^{(i)} \ \dots \ u_{k^{(1)}-1}^{(i)}) \cdot \G^{(10)} \!
 = \u^{(i+1)}  \G^{(0)} + \e^{(i+1)}\label{eq:substract_decoding}\\
 &\r^{(i-1)}\! - (u_0^{(i-1)} \ u_1^{(i-1)} \ \dots \ u_{k^{(1)}-1}^{(i-1)})\cdot \G^{(00)}\nonumber\\
 &\hspace{1ex}=(u_{k^{(1)}}^{(i-1)}\dots \ u_{k-1}^{(i-1)} \ | \ u_0^{(i-2)}\dots \ u_{k^{(1)}-1}^{(i-2)})  \begin{pmatrix}\mathbf G_{01}\\\mathbf G_{10}\end{pmatrix} \!+ \e^{(i-1)}.\nonumber
\end{align}
Hence, Step~2 uses the information from block $i$ to
decode $\pumforward{i}$ blocks forward with $\BMD{\mycode{C}_0}$ and $\pumbackward{i}$ blocks backward with $\BMD{\mycode{C}_1} $
from \emph{any} node found in Step~1. This closes (most of) the gaps between two blocks \emph{correctly} decoded by $\BMD{\Ca}$ (of course, it is not known, which blocks are decoded correctly).

We define the values $\pumforward{i}$ and $\pumbackward{i}$ as follows:
\begin{small}
\begin{align}
\pumforward{i}&\!=\!\min_{j}\!\Bigg(\!j\Big\arrowvert\!\sum_{h=1}^{j}\!\big(\da-\!m^{(i+h)}\big)\!\geq\! \frac{\dcdes{j}-\!\sum\limits_{h=1}^{j}\!\!\left(\numbRowErasures^{(i+h)} + \numbColErasures^{(i+h)}\right)}{2}\Bigg),\label{eq:deflforward}\\
\pumbackward{i}\!&\!= \min_{j}\!\Bigg(\!j\Big\arrowvert\!\sum_{h=1}^j\!\big(\da-\!m^{(i-h)}\big)\!\geq\! \frac{\drcdes{j}-\!\sum\limits_{h=1}^{j}\!\!\left(\numbRowErasures^{(i-h)} + \numbColErasures^{(i-h)}\right)}{2}\Bigg). \label{eq:deflbackward}
\end{align}
\end{small}
These definitions are chosen such that we can guarantee correct decoding if the BRD condition \eqref{eq:bmdcond} is fulfilled (see Section~\ref{subsec:proofdecoding}).

For Step~3 and some $i \in \intervallexcl{0}{N}$, assume we know $(u_0^{(i+1)} \ u_1^{(i+1)} \ \dots \ u_{k^{(1)}-1}^{(i+1)})$ and $\u^{(i)}$ from Step~1 or 2, then as in \eqref{eq:substract_decoding}, we can calculate
\begin{align*}
\r^{(i+1)}\! &-\!  (u_0^{(i+1)}  \dots \ u_{k^{(1)}-1}^{(i+1)})\!\cdot\!\G^{(00)}\!- (u_0^{(i)}  \dots \ u_{k^{(1)}-1}^{(i)}) \!\cdot\! \mathbf G^{(10)} \\ & 
\hspace{5ex}= (u_{k^{(1)}}^{(i+1)} \ u_{k^{(1)}+1}^{(i+1)} \ \dots \ u_{k^{(1)}-1}^{(i+1)}) \cdot \mathbf G^{(01)} + \e^{(i+1)},
\end{align*}
which shows that we can
use $\BMD{\mycode{C}_{01}}$ to close a remaining gap in block $i+1$. 

After Step~3, assign as metric to each edge

\begin{small}
\begin{align}
\hspace{-8ex}m^{(i)}=\!\begin{cases}
       \ \rk(\r^{(i)}-\c^{(i)\prime}), &\parbox[t]{0.3\linewidth}{\text{if \BMD{\mycode{C}_0}, \BMD{\mycode{C}_1} or} \text{\BMD{\mycode{C}_{01}} successful,}}\\[2ex]
		  \bigg\lfloor\dfrac{d_{01}+1+\numbRowErasures^{(i)}+\numbColErasures^{(i)}}{2}\bigg\rfloor, \quad \qquad& \text{else},
                 \end{cases} \nonumber\\[-4ex]
          \label{eq:defmetric_step3}
\end{align}
\end{small}
\hspace{-1ex}$\forall i \in \intervallincl{0}{N}$, where $\c^{(i)\prime}$ denotes the result of a successful decoding. 
For one received block $\r^{(i)}$, there can be several decoding results $\c^{(i)\prime}$ from the different BMD decoders.
Thus, there can be more than one edge in the reduced trellis at depth $i$. Each edge is labeled with regard to \eqref{eq:defmetric_step3} using its corresponding code block.

\begin{figure*}[ht]
\centering
\scalebox{0.8}{\includegraphics{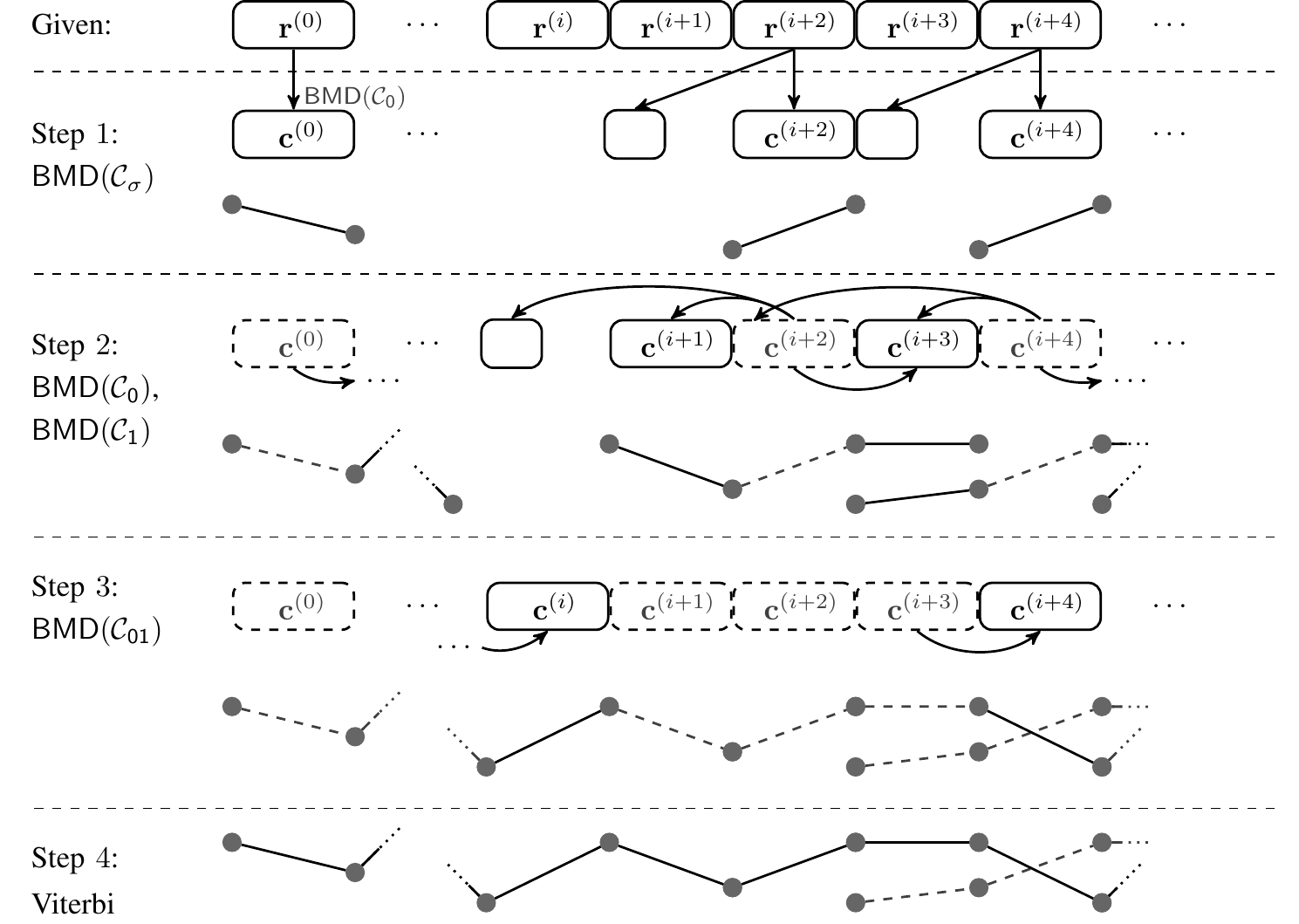}} 
\caption{Illustration of the different steps of Algorithm~\ref{algo:pum}: The received sequence $(\r^{(0)} \ \r^{(1)} \ \dots \ \r^{(N)})$ is given and the different steps and their decoding results are shown. Dashed blocks/edges illustrate that they were found in a previous step.}
\label{fig:decoding_pum}
\end{figure*}

Finally, we use the Viterbi algorithm to find the path of smallest sum rank weight in this reduced trellis. 
As in \cite{DettmarSorger_BMDofUM}, we use $m^{(i)}$, for all $i\in \intervallincl{0}{N}$, as edge metric and the sum over different edges as path metric. 
The different steps of our decoding algorithm are roughly summarized in Algorithm~\ref{algo:pum}, the details can be found in the preceding description and Figure~\ref{fig:decoding_pum} illustrates our decoding algorithm.

In Section~\ref{subsec:proofdecoding}, we prove that if \eqref{eq:bmdcond} is fulfilled, then after the three block decoders, all gaps are closed
and the Viterbi algorithm finds the path with the smallest sum rank weight.

\subsection{Proof of Correctness}\label{subsec:proofdecoding}
In the following, we prove that decoding with Algorithm~\ref{algo:pum} is successful if the BRD condition \eqref{eq:bmdcond} is fulfilled.
The proof follows the proof of Dettmar and Sorger \cite{Dettmar_PUM_PhD_1994,DettmarSorger_BMDofUM}.
Lemma~\ref{lem:gapnottoobig} shows that the gaps between two \emph{correct} results of Step~1 are not too big and
Lemmas~\ref{lem:corrpathlesserr} and~\ref{lem:wintergap} show that the gap size after Steps~1 and 2 
is at most one if the BRD condition \eqref{eq:bmdcond} is fulfilled. Theorem~\ref{thm:gapsclosed} shows that these
gaps can be closed with $\BMD{\mycode{C}_{01}}$ and the Viterbi algorithm finds the correct path.

\begin{lemma}[Gap Between two Correct Results of Step~1]\label{lem:gapnottoobig}
If the BRD condition \eqref{eq:bmdcond} is satisfied, then 
the length of any gap between two correct decisions in Step~1 of Algorithm~\ref{algo:pum}, denoted by $\c^{(i)}$, $\c^{(i+j)}$, is less than $\min\{ \pumforwardBig{i},\pumbackwardBig{i}\}$, where\\[-5ex]

\begin{small}
\begin{align*}
\pumforwardBig{i}&\!=\!\min_{j}\!\Bigg(\!j\big\arrowvert\!\sum\limits_{h=1}^{j} \!\big(\da-m^{(i+h)}\big)\! \geq \!\! \frac{\drdes{j}-\sum\limits_{h=1}^{j}(\numbRowErasures^{(i+h)} + \numbColErasures^{(i+h)})}{2}\!\Bigg),\\
\pumbackwardBig{i}&\!=\!\min_{j}\!\Bigg(\!j\big\arrowvert\!\sum\limits_{h=1}^{j} \!\big(\da-m^{(i-h)}\big)\! \geq \!\! \frac{\drdes{j}-\sum\limits_{h=1}^{j}(\numbRowErasures^{(i-h)} + \numbColErasures^{(i-h)})}{2}\!\Bigg).
\end{align*}
\end{small}
\end{lemma}

\begin{IEEEproof}
Decoding of a block $\r^{(i)}$ in Step~1 fails or outputs a wrong result if there are 
at least $(\da-\numbRowErasures^{(i)}-\numbColErasures^{(i)})/2$ errors in rank metric. 
In such a case, the metric $m^{(i)} =\lfloor (\da+1+\numbRowErasures^{(i)}+\numbColErasures^{(i)})/2\rfloor $ is assigned. 

In order to prove the statement, assume there is a gap of at least $\pumforwardBig{i}$ blocks after Step~1. 
Then,
\begin{align*}
 \sum\limits_{h=1}^{\pumforwardBig{i}}t^{(i+h)}
 &\geq \sum\limits_{h=1}^{\pumforwardBig{i}}\frac{\da-\numbRowErasures^{(i+h)}-\numbColErasures^{(i+h)}}{2}
\geq\sum\limits_{h=1}^{\pumforwardBig{i}}\! \big(\da-m^{(i+h)}\big)
\\&\geq
\frac{\drdes{\pumforwardBig{i}}-\sum\limits_{h=1}^{\pumforwardBig{i}}\big(\numbRowErasures^{(i+h)}+\numbColErasures^{(i+h)}\big)}{2},
\end{align*}
which follows from the definition of the metric \eqref{eq:defmetric} and from the definition of $\pumforwardBig{i}$. 
This contradicts the BRD condition \eqref{eq:bmdcond}. 
Similarly, we can prove this for $\pumbackwardBig{i+j}$ and hence, the gap size has to be less than $\min\{ \pumforwardBig{i},\pumbackwardBig{i}\}$.
\end{IEEEproof}
Note that $\pumforwardBig{i}$ and $\pumforward{i}$ differ only in using the active row rank distance $\drdes{j}$ and the active column rank distance $\dcdes{j}$, respectively. 
Further, Lemma~\ref{lem:gapnottoobig} will not be used in the following, but it shows an upper bound on the size of the gaps between two correctly decoded blocks after the first step of our decoding algorithm.

\begin{lemma}[Correct Path for Few Errors]\label{lem:corrpathlesserr}
Let $\c^{(i)}$ and $\c^{(i+j)}$ be decoded correctly in Step~1 of Algorithm~\ref{algo:pum}. 
Let Step~2 of Algorithm~\ref{algo:pum} decode $\pumforward{i}$ blocks in forward direction starting in $\c^{(i)}$, and $\pumbackward{i+j}$ blocks in backward direction starting in $\c^{(i+j)}$ (see also \eqref{eq:deflforward}, \eqref{eq:deflbackward}).

Then, the correct path is in the reduced trellis if the BRD condition \eqref{eq:bmdcond} is satisfied and if in each block
less than $\min \lbrace (d_0-\numbRowErasures^{(i)}-\numbColErasures^{(i)})/2, (d_1-\numbRowErasures^{(i)}-\numbColErasures^{(i)})/2 \rbrace$ rank
errors occurred.
\end{lemma}
\begin{IEEEproof}
If there are 
less than $\min \left\lbrace (d_0-\numbRowErasures^{(i)}-\numbColErasures^{(i)})/2, (d_1-\numbRowErasures^{(i)}-\numbColErasures^{(i)})/2 \right\rbrace$
errors in a block, \BMD{\mycode{C}_0} and \BMD{\mycode{C}_1} always yield the correct decision.
Due to the definition of $\pumforward{i}$, see \eqref{eq:deflforward}, the forward decoding with \BMD{\mycode{C}_0} terminates as soon as 
\begin{align*}
\sum\limits_{h=1}^{\pumforward{i}} t^{(i+h)}
 &\geq \sum\limits_{h=1}^{\pumforward{i}}\frac{\da-\numbRowErasures^{(i+h)}-\numbColErasures^{(i+h)}}{2}
\geq \sum\limits_{h=1}^{\pumforward{i}}\big(\da-m^{(i+h)}\big)\\
&\geq \frac{\dcdes{\pumforward{i}}-\sum\limits_{h=1}^{\pumforward{i}}\big(\numbRowErasures^{(i+h)}+\numbColErasures^{(i+h)}\big)}{2}\nonumber
\\&=\frac{d_0}{2} + \big(\pumforward{i}-1\big)\frac{\da}{2}-\frac{\sum_{h=1}^{\pumforward{i}}\big(\numbRowErasures^{(i+h)}+\numbColErasures^{(i+h)}\big)}{2},
\end{align*}
where the first inequality holds since the decoding result could not be found in Step~1 and the second and third hold due to the definition of the metric \eqref{eq:defmetric} and the definition of $\pumforward{i}$.

Similarly, backward decoding with \BMD{\mycode{C}_1} terminates if 
\begin{align*}
\sum\limits_{h=1}^{\pumbackward{i+j}} t^{(i+j-h)} 
\geq &
\frac{d_1}{2} + \big(\pumbackward{i+j}-1\big)\frac{\da}{2}\\
&\hspace{2ex}-\frac{\sum_{h=1}^{\pumbackward{i+j}}\big(\numbRowErasures^{(i+j-h)}+\numbColErasures^{(i+j-h)}\big)}{2}.
\end{align*}
The correct path is in the reduced trellis if $\pumforward{i}+\pumbackward{i+j}\geq j-1$, since the gap is then closed.
Assume now on the contrary that $\pumforward{i}+\pumbackward{i+j}<j-1$.
Since Step~1 was not successful for the blocks in the gap, at least $(\da-\numbRowErasures^{(h)}-\numbColErasures^{(h)})/2$ rank errors occured in every block $\r^{(h)}$, $\forall h \in \intervallexcl{i+\pumforward{i}+1}{i+j-\pumbackward{i+j}}$, i.e, in the blocks in the gap between the forward and the backward path.
Then,
\begin{align*}
&\sum\limits_{h=1}^{j-1} t^{(i+h)}
\geq \frac{\dcdes{\pumforward{i}}-\sum_{h=1}^{\pumforward{i}}(\numbRowErasures^{(i+h)}+\numbColErasures^{(i+h)})}{2}\\
 &\geq \sum\limits_{h=1}^{\pumforward{i}} t^{(i+h)}\! +\! \sum\limits_{h=1}^{\pumbackward{i+j}}\! t^{(i+j-h)} 
\!+\!\!\sum\limits_{h=\pumforward{i}+1}^{j-1-\pumbackward{i+j}}\!\!\!\frac{\da-\numbRowErasures^{(i+h)}+\numbColErasures^{(i+h)}}{2}\\
&\geq \frac{d_0}{2} + \big(\pumforward{i}-1\big)\frac{\da}{2}-\frac{\sum\limits_{h=1}^{\pumforward{i}}(\numbRowErasures^{(i+h)}+\numbColErasures^{(i+h)})}{2}\\
&\hspace{2ex}+\frac{d_1}{2} + \big(\pumbackward{i+j}-1\big)\frac{\da}{2}-\frac{\sum\limits_{h=1}^{\pumbackward{i+j}}\big(\numbRowErasures^{(i+j-h)}+\numbColErasures^{(i+j-h)}\big)}{2}\\
&\hspace{2ex}+\frac{\big(j-1-\pumforward{i}-\pumbackward{i+j}\big)}{2}\da-\frac{\!\!\sum\limits_{h=\pumforward{i}+1}^{j-1-\pumbackward{i+j}}\!\!\!\big(\numbRowErasures^{(i+h)}+\numbColErasures^{(i+h)}\big)}{2}\\
&\geq \frac{d_0}{2} +\frac{d_1}{2}+\frac{(j-3)}{2}\da- \frac{\sum_{h=1}^{j-1}\big(\numbRowErasures^{(i+h)}+\numbColErasures^{(i+h)}\big)}{2}\\
&=\frac{\drdes{j-1}-\sum_{h=1}^{j-1}\big(\numbRowErasures^{(i+h)}+\numbColErasures^{(i+h)}\big)}{2},
\end{align*}
which is a contradiction to the BRD condition \eqref{eq:bmdcond} and the statement follows. 
\end{IEEEproof}

\begin{lemma}[Gap Size is at Most One After Steps~1 and 2]\label{lem:wintergap}
Let $\c^{(i)}$ and $\c^{(i+j)}$ be decoded correctly in Step~1 of Algorithm~\ref{algo:pum} (with no other correct decisions in between) and let the BRD condition
\eqref{eq:bmdcond} be fulfilled. 
Let $d_0 = d_1$.

Then, there is at most one error block $\e^{(h)}$, $h \in \intervallexcl{i+1}{i+j}$, of rank at least $ (d_0-\numbRowErasures^{(i)}-\numbColErasures^{(i)})/2$. 
\end{lemma}
\begin{IEEEproof}
To fail in Step~1, there have to be at least $(\da-\numbRowErasures^{(i)}-\numbColErasures^{(i)})/2$ errors in $\r^{(i)}$, $\forall i \in \intervallexcl{i+1}{i+j}$.
If two error blocks in this gap have rank at least $ (d_0-\numbRowErasures^{(i)}-\numbColErasures^{(i)})/2$, then 
\begin{align*}
\sum\limits_{h=1}^{j-1}t^{(i+h)}
&\geq2\cdot\frac{d_0}{2}+(j-3)\cdot \frac{\da}{2}
-\frac{\sum\limits_{h=1}^{j-1}\big(\numbRowErasures^{(i+h)}+\numbColErasures^{(i+h)}\big)}{2}\\
&\geq \frac{\drdes{j-1}}{2}-\frac{\sum_{h=1}^{j-1}\big(\numbRowErasures^{(i+h)}+\numbColErasures^{(i+h)}\big)}{2},
\end{align*}
which contradicts \eqref{eq:bmdcond}.
\end{IEEEproof}
Lemmas~\ref{lem:corrpathlesserr} and~\ref{lem:wintergap} show that if the BRD condition is satisfied, then the correct path is in the reduced trellis after Steps 1 and 2, except for at most one block.

\begin{theorem}[Correct Path is in Reduced Trellis]\label{thm:gapsclosed}
If the BRD condition \eqref{eq:bmdcond} is satisfied, then the correct path is in the reduced trellis after Step~3 of Algorithm~\ref{algo:pum}.
\end{theorem}
\begin{IEEEproof}
Lemmas~\ref{lem:corrpathlesserr} and~\ref{lem:wintergap} guarantee that after Step~2, at most one block of the correct path is missing in the reduced trellis.
For one block, say $\r^{(h)}$, it follows from the BRD condition that $(2 \cdot t^{(h)} + \numbRowErasures^{(h)} + \numbColErasures^{(h)}) <\drdes{1}=\dfree$ and any decoder of distance at least $\drdes{1}$ is able to decode correctly in this block.
Hence, after Step~3, $\BMD{\mycode{C}_{01}}$ is able to find the correct solution for this block since $d_{01} = \drdes{1} $ and the correct path is in the reduced trellis. 
\end{IEEEproof}

The complexity is determined by the complexity of the BMD rank block error-erasure decoders from Table~\ref{tab:subcodes_pum}, 
which are all in the order $\OCompl{n^2}$ operations in $\F$.
Hence, the calculation of the complexity of Algorithm~\ref{algo:pum} is straight-forward to \cite[Theorem~3]{DettmarSorger_BMDofUM} and we can give the following bound on the complexity without proof.

\begin{theorem}[BRD Decoding with Algorithm~\ref{algo:pum}]\label{thm:pum_decoding}
Let $k + k^{(1)} \leq n \leq m$, where $k^{(1)} \leq k$.
Let $\mycode{C}$ be a zero-forced terminated $\UM{n,k}$ or $\PUM{n,k}{k^{(1)}}$ code over $\F$ as in Definition~\ref{def:pumgab_genmat}.
Let a received sequence $\r = (\r^{(0)} \ \r^{(1)}\ \dots \ \r^{(N)}) \in \F^{n(N+1)}$ be given.

Then, Algorithm~\ref{algo:pum} finds the code sequence $\c = (\c^{(0)} \ \c^{(1)}\ \dots \ \c^{(N)}) \in \F^{n(N+1)}$ with smallest sum rank distance to $\r$ if the BRD condition from~\eqref{eq:bmdcond} is satisfied. The complexity of decoding one block of length $n$ is at most
$\OCompl{\da n^2} \leq \OCompl{n^3}$
operations in $\F$.
\end{theorem}


\subsection{Decoding of the Arbitrary-Rate Code Construction}
For the arbitary-rate code construction from Section~\ref{subsec:arb-rate-constr}, our decoding algorithm from the previous section can be modified straight-forward to \cite{WachterStinner_BMDArbitraryRates_2012_conf}. Hence, we outline this adaption only shortly here and refer the reader to \cite{WachterStinner_BMDArbitraryRates_2012_conf} for details.

The linear dependencies in the matrices $\G^{(0)\prime}$ and $\G^{(1)\prime}$ 
(see Definition~\ref{def:pumarbitrary}) have the effect that $\ell$ consecutive zero blocks within the code sequence are possible (compare Section~\ref{subsec:arb-rate-constr}).
Further, the dependencies spread the information to $\ell+1$ blocks
and we can therefore guarantee to reconstruct a certain information block $\u^{(i)}$ only if $\ell+1$ consecutive blocks (including code block $\u^{(i)}$) could be decoded.
This is shown in the following example.

\begin{example}[Reconstructing Information Block]\label{ex:reconst_info}
Let $\varphi = 2 k^{(1)}/3$, where $\ell=2$ and $\mathbf \Phi$ has 
twice as much rows as $\A$. Assume, we have decoded $\c^{(0)}$, $\c^{(1)}$ and $\c^{(2)}$ and we want to reconstruct $\u^{(1)}$. 

We decompose $\u^{(0)}, \u^{(1)}, \u^{(2)}$ into sub-blocks, i.e.:
$\u^{(j)} = (\u^{(j)}_1\; |\; \u^{(j)}_2\; |\;\u^{(j)}_3\; |\;\u^{(j)}_4 )$ for $j=0,1,2$, where the first three sub-blocks have length $k^{(1)}-\varphi$ and the last sub-block has length $k-k^{(1)}$. Then,
\begin{align*}
\c^{(1)} &= 
(\u^{(1)}_1 |\; \u^{(1)}_2 + \u^{(0)}_1  |\;\u^{(1)}_3 + \u^{(0)}_2  |\;\u^{(1)}_4  |\;\u^{(0)}_3 )
\!\cdot\!
\begin{pmatrix}
\A \\
\mathbf{\Phi}_1\\
\mathbf{\Phi}_2\\
\G^{(01)\prime}\\
\B
\end{pmatrix}\\
&\defeq \widehat{\u}^{(1)} \cdot \G_{all}^\prime,
\end{align*}
where $\mathbf \Phi = \left(\begin{smallmatrix}
\mathbf \Phi_1\\\mathbf \Phi_2
\end{smallmatrix}\right)$ and $\mathbf \Phi_1$, $\mathbf \Phi_2$ each have $k^{(1)}-\varphi$ rows. 
Since we know $\c^{(1)}$, and since $\G_{all}^\prime$ defines an MRD code, we can reconstruct the vector $\widehat{\u}^{(1)}$. 
This directly gives us $\u_1^{(1)}$ and $\u_4^{(1)} $. This reconstruction can be done in the same way for $\c^{(0)}$ and we obtain (amongst others) $\u_1^{(0)}$. 
To obtain $\u_2^{(1)}$, we subtract $\u_1^{(0)}$ from the known sum $\u_2^{(1)} +\u_1^{(0)}$.
The reconstruction for $\c^{(2)}$ provides $\u_3^{(1)} $ and we have recovered the whole information block $\u^{(1)}$. 

This example has shown why $\ell+1$ consecutive decoded blocks are necessary to reconstruct one information block. 
It does not matter if the other decoded blocks precede or succeed the required information block.
\end{example}

Apart from the reconstruction of the information, there are further parts in the decoding algorithm which have to be modified.
An error of \emph{minimum weight} causing a sequence of non-reconstructible information blocks in the first decoding step has the following structure:
\begin{align*}
(\underbrace{0,\dots,0,\times}_{\ell+1 \text{ blocks}} \,|\,\underbrace{0,\dots,0,\times}_{\ell+1 \text{ blocks}}\,|\,\dots\,|\,\underbrace{0,\dots,0,\times}_{\ell+1 \text{ blocks}}
\,|\,\underbrace{0,\dots,0}_{\ell \text{ blocks}}),
\end{align*}
where $\times$ marks blocks (of length $n$) of rank weight at least $\da/2$.
In this case, also the information of the $\ell$ error-free blocks between the erroneous blocks cannot be reconstructed since we need $\ell+1$ consecutive decoded blocks to reconstruct the information. Further, the last $\ell$ error-free blocks make it necessary to decode $\ell$ additional steps in forward direction in the second step of Algorithm~\ref{algo:pum}.

In order to decode with Algorithm~\ref{algo:pum}, we have to take into account the slower increase of the resulting extended distances due to the sequences of possible zero code blocks.
Hence, as in \cite{WachterStinner_BMDArbitraryRates_2012_conf}, we generalize \eqref{eq:deflforward} by simply subtracting $\ell$ in the summation, which is equivalent to going $\ell$ steps further:

\begin{small}
\begin{equation}
\pumforwardprime{i}\!=\!\min_{j}\!\Bigg(\!j\Big\arrowvert\!\sum_{h=1}^{j-\ell}\!\frac{\da^\prime-\!m^{(i+h)}}{\ell+1}\!\geq\! \frac{\dcdesprime{j}-\!\sum\limits_{h=1}^{j}\!\!\left(\numbRowErasures^{(i+h)} + \numbColErasures^{(i+h)}\right)}{2}\Bigg),\label{eq:deflforwardarb}
\end{equation}
\end{small} 
\hspace{-1ex}which reduces to \eqref{eq:deflforward} for $\ell=0$. Further $\pumbackwardprime{i} = \pumbackward{i}$. 

Hence, in order to decode the arbitrary-rate construction we have to modify Algorithm~\ref{algo:pum} as follows:
\begin{itemize}
\item the reconstruction of information blocks requires $\ell+1$ consecutive code blocks as in Example~\ref{ex:reconst_info},
\item the path extension $\pumforward{i}$ has to be prolonged as in~\eqref{eq:deflforwardarb},
\item the metric definitions in \eqref{eq:defmetric}, \eqref{eq:defmetric_step3} have to be modified by simply adding "and $\u^{(i)}$ could be reconstructed" in the if-part of both definitions.
\end{itemize}

Then, as in Section~\ref{subsec:proofdecoding}, we can guarantee that the correct path is in the reduced trellis if
\begin{align*}
&\sum\limits_{h=i}^{i+j-1}\Big(2 \cdot t^{(h)} + \numbRowErasures^{(h)} + \numbColErasures^{(h)}\Big) <\drdesprime{j} , \\
&\qquad \forall i \in \intervallincl{0}{N}, j \in \intervallincl{0}{N-i+1},\nonumber
\end{align*}
where $\drdesprime{j}$ is defined as in~\eqref{eq:ext_dist_arb_rate}.

\section{Application to Random Linear Network Coding}\label{sec:pum_networkcoding}
Our motivation for considering convolutional codes in rank metric is to apply them in multi-shot \emph{random linear network coding} (RLNC).
In this section, we first explain the model of multi-shot network coding and show how to define lifted (P)UM codes in rank metric.
Afterwards, we show how decoding of these lifted (P)UM codes reduces to error-erasure decoding of (P)UM codes in rank metric.


\subsection{Multi-Shot Transmission of Lifted PUM Codes}
As network channel model we assume a \emph{multi-shot} transmission over the so-called \emph{operator channel}. The operator channel was defined by K\"otter and Kschischang in \cite{koetter_kschischang} and the concept of multi-shot transmission over the operator channel was first considered by Nóbrega and Uch\^{o}a-Filho \cite{NobregaUchoa-2010Multishot}.

In this network model, a source transmits packets (which are vectors over a finite field) to a sink. The network has several directed links between the source, some internal nodes and the sink.
The source and sink apply coding techniques for error control, but have no knowledge about the structure of the
network. This means, we consider \emph{non-coherent} RLNC. 
In a multi-shot transmission, we use the network several times and the internal structure may change 
in every time instance. In detail, we assume that we use it $N+1$ times. In the following, we shortly give basic notations for this network channel model.
The notations are similar to \cite{silva_rank_metric_approach}, but we include additionally the time dependency.

Let $\mathbf X^{(i)}\in \Fq^{n \times (n+m)}$, $\forall i\in \intervallincl{0}{N}$. 
The rows represent the transmitted packets $X^{(i)}_0,X^{(i)}_1,\dots,X^{(i)}_{n-1}$ $ \in \Fq^{n+m}$ at time instance (shot) $i$. Similarly, let 
$\mathbf Y^{(i)} \in \Fq^{n^{(i)}\times (n+m)}$ be a matrix whose $n^{(i)}$ rows correspond to 
the received packets $Y^{(i)}_0,Y^{(i)}_1,\dots,Y^{(i)}_{n^{(i)}-1}\in \Fq^{m+n}$.
Notice that $n$ and $n^{(i)}$ do not have to be equal since packets can be erased and/or additional packets might be inserted.

The term \emph{random linear network coding} originates from the behavior of the internal nodes:
they create random linear combinations of the 
packets received so far in the current shot $i$, $\forall i\in \intervallincl{0}{N}$. Additionally, erroneous packets might be inserted into the network and transmitted packets might be lost or erased.

Let the links in the network be indexed from $0$ to $\ell-1$, then, as in \cite{silva_rank_metric_approach}, 
let the rows of a matrix $\mathbf Z^{(i)} \in \Fq^{\ell \times (n+m)}$ contain the error packets $Z^{(i)}_0,Z^{(i)}_1,\dots, Z^{(i)}_{\ell-1}$ 
inserted at the links $0$ to $\ell-1$ at shot $i$. If $Z^{(i)}_j = 0$, $j \in \intervallexcl{0}{\ell}$, then no corrupt packet was inserted at link $j \in \intervallexcl{0}{\ell}$ and time $i$.
Due to the linearity of the network, the output can be written as:
\begin{equation}\label{eq:operatorchannel}
\Y^{(i)} = \A^{(i)} \X^{(i)} + \B^{(i)} \Z^{(i)},
\end{equation}
where $\mathbf A^{(i)} \in \Fq^{n^{(i)} \times n}$ and $\mathbf B^{(i)} \in \Fq^{n^{(i)} \times \ell}$ are the (unknown) channel transfer matrices 
at time $i$. 


When there are no errors or erasures in the network, the row space of $\Y^{(i)}$ is the same as the row space of $\X^{(i)}$.
In \cite{koetter_kschischang,silva_rank_metric_approach} it was shown that subspace codes constructed by lifted MRD codes (as in Lemma~\ref{lem:subspacecode_lifted_mrd}) provide an almost optimal solution to error control in the operator channel.
Such lifted MRD codes are a special class of constant-dimension codes (see Lemma~\ref{lem:subspacecode_lifted_mrd}).
In the following, we define \emph{lifted PUM codes} based on Gabidulin codes in order to use these constant-dimension codes for error correction in multi-shot network coding.

\begin{definition}[Lifted (Partial) Unit Memory Code]\label{def:liftedpum}
Let $\mycode{C}$ be a zero-forced terminated $\UM{n,k}$ or $\PUM{n,k}{k^{(1)}}$ code over $\F$ as in Definition~\ref{def:pumgab_genmat}. 
Represent each code block $\c^{(i)} \in \F^n$, $\forall i \in \intervallincl{0}{N}$, as matrix $\C^{(i)}=\extsmallfieldinput{\c^{(i)}} \in \Fq^{m\times n}$. 

Then, the lifting of $\mycode{C}$ is defined by the following set of subspace sequences:
\begin{align*}
\liftmap{\mycode{C}} \defeq \Big\lbrace \Big( \RowspaceNoInput \big( [\I_n \ & \C^{(0)T}]\big)  \ \dots \ \RowspaceNoInput \big( [\I_n \ \C^{(N)T}]\big) \Big)  :\\
&\Big(\extsmallfieldInverse(\C^{(0)}) \ \dots \ \extsmallfieldInverse(\C^{(N)}) \Big) \in \mycode{C} \Big\rbrace.
\end{align*}
\end{definition}
As in Definition~\ref{def:lifting_matrixcode}, we denote $\liftmap{\C^{(i)T}} = \RowspaceNoInput \big( [\I_n \ \C^{(i)T}]\big)$, $\forall i\in \intervallincl{0}{N}$.
We transmit this sequence of subspaces over the operator channel such that each transmitted matrix is a lifted block of a codeword of the rank-metric PUM code, i.e.,
 $\X^{(i)} = [\I_n \ \C^{(i)T} ]$, $\forall i \in \intervallincl{0}{N}$. Of course, any other basis of the row space can also be chosen as transmitted matrix.
 
By means of this lifted PUM code, we create dependencies between the different shots in the network. 
Since each code block of length $n$ is a codeword of the block code $\Ca$, each transmitted subspace is a codeword of a $\CDC{n+m,\SubspacedistNoInput=2\da,n}$ constant-dimension code, lying in $\Grassm{n+m,n}$, see \cite[Proposition~4]{silva_rank_metric_approach} and Lemma~\ref{lem:subspacecode_lifted_mrd}.
However, the lifted (P)UM code contains additionally dependencies between the different blocks and for decoding, we obtain therefore a better performance than simply lifting the \emph{block} code $\Ca$ as in Lemma~\ref{lem:subspacecode_lifted_mrd}. Since the PUM code transmits $k$ information symbols per shot, a comparison with a lifted block code of rate $k/n$ is much fairer than comparing it with $\Ca$ (see also Example~\ref{ex:decoding}).

\subsection{Decoding of Lifted PUM Codes in the Operator Channel}
In this section, we will show how the decoding problem in the operator channel reduces to error-erasure decoding of PUM codes based on Gabidulin codes---analog to \cite{silva_rank_metric_approach}, where it reduces to error-erasure decoding of Gabidulin codes. 
Since each code block of length $n$ of a $\PUM{n,k}{k^{(1)}}$ code is a codeword of the block code $\Ca$, we can directly use the reformulations of Silva, Kschischang and Kötter \cite{silva_rank_metric_approach}.

Let the transmitted matrix at time instance $i$ be $\X^{(i)} = [\I_n \ \C^{(i)T} ]$ and denote by 
$\Y^{(i)} = [\widehat{\mathbf A}^{(i)}\ \widehat{\mathbf Y}^{(i)}] \in \Fq^{n^{(i)} \times (n+m)}$ the received matrix 
after the multi-shot transmission over the operator channel as in \eqref{eq:operatorchannel}. 
The channel transfer matrices $\A^{(i)}$ and $\B^{(i)}$ are time-variant. 
Moreover, assume $\rank(\Y^{(i)}) = n^{(i)}$, since linearly dependent received packets are directly discarded.
Then, as in \cite{silva_rank_metric_approach}, we denote the column and row deficiency of $\widehat{\A}^{(i)}$ by:
\begin{equation*}
\numbColErasures^{(i)} \defeq n - \rank(\widehat{\A}^{(i)}), \ \numbRowErasures^{(i)} \defeq n^{(i)} - \rank(\widehat{\A}^{(i)}), \ \forall i \in \intervallincl{0}{N}.
\end{equation*}
If we calculate the \emph{reduced row echelon} (RRE) form of $\Y^{(i)}$ (and fill it up with zero rows, if necessary), we obtain the following matrix in $\Fq^{(n+\numbRowErasures^{(i)})\times(n+m)}$ (similar to \cite[Proposition~7]{silva_rank_metric_approach}, but in our notation):
\begin{equation} \label{eq:rre_network}
\RRE_0\big(\Y^{(i)}\big) = 
\begin{pmatrix}
\mathbf I_n + \B^{(i,C) T} \mathbf I_{\myset{U}^{(i)}}^T & \R^{(i)T}\\
\0 & \A^{(i,R)T}
\end{pmatrix},
\end{equation}
for a set $\myset{U}^{(i)} \subseteq \intervallincl{0}{n-1}$ with $|\myset{U}^{(i)}| = \numbColErasures^{(i)}$ such that
$\I_{\myset{U}^{(i)}}^T \R^{(i)T} = \0$ and $\I_{\myset{U}^{(i)}}^T \B^{(i,C) T} = - \I_{\numbColErasures^{(i)}}$,
and $\I_{\myset{U}^{(i)}}$ denotes the submatrix of $\I_n$ consisting of the columns indexed by $\myset{U}^{(i)}$. 
Moreover, $\B^{(i,C) T}\in \Fq^{n \times \numbColErasures^{(i)}}$ and $\A^{(i,R)T} \in \Fq^{\numbRowErasures^{(i)}\times n}$.

Furthermore, it was shown in \cite{silva_rank_metric_approach} that $\R^{(i)}$ can be decomposed into
\begin{equation*}
\R^{(i)} = \C^{(i)} +  \A^{(i,R)} \B^{(i,R)} + \A^{(i,C)} \B^{(i,C)} + \A^{(i,E)} \B^{(i,E)}, 
\end{equation*}
$\forall i \in \intervallincl{0}{N}$, where $\big(\extsmallfieldInverse(\C^{(0)})  \ \dots \ \extsmallfieldInverse(\C^{(N)}) \big) \in \mycode{C}$ and $\A^{(i,R)}$ and $\B^{(i,C)}$ are known to the receiver, since the matrix from~\eqref{eq:rre_network} can be calculated from the channel output.
Comparing this equation to \eqref{eq:decomp_errrorerasures} makes clear that
the problem of decoding lifted PUM codes (as in Definition~\ref{def:liftedpum}) in the operator channel reduces to error-erasure decoding of the PUM code in rank metric.
For this purpose, we can use our decoding algorithm from Section~\ref{sec:pum_decoding},
which is based on rank-metric error-erasure \emph{block} decoders. 

Now, let the received matrix sequence $\Y = (\Y^{(0)} \ \Y^{(1)} \ \dots \ \Y^{(N)})$ as output of the operator channel 
be given, then Algorithm~\ref{algo:dec_network} shows how to reconstruct the transmitted information sequence.

\printalgoIEEEWidth{ \vspace{1ex}
 \caption{\newline$\u  \leftarrow$ \textsc{NetworkPUMDecoder}$\big(\Y\big)$}
\label{algo:dec_network}
 \SetKwInput{KwIn}{\underline{Input}}
 \SetKwInput{KwOut}{\underline{Output}}
 \SetKwInput{KwIni}{\underline{Initialize}}
\DontPrintSemicolon
\SetAlgoVlined
\LinesNumbered
\KwIn{Received sequence $\mathbf{Y}= (\Y^{(0)} \  \Y^{(1)} \ \dots, \Y^{(N)})$,\newline where $\Y^{(i)} \in \Fq^{n^{(i)} \times (n+m)}$, $\forall i \in \intervallincl{0}{N}$}
\BlankLine
$\numbColErasures^{(i)}\leftarrow n - \rank(\widehat{\A}^{(i)})$, $\forall i \in \intervallincl{0}{N}$\;
$\numbRowErasures^{(i)} \leftarrow n^{(i)} - \rank(\widehat{\A}^{(i)})$, $\forall i \in \intervallincl{0}{N}$
\BlankLine
Calculate $\RRE_0(\Y^{(i)})$ and $\R^{(i)}$ as in \eqref{eq:rre_network}, $\forall i \in \intervallincl{0}{N}$\;
\BlankLine
$\mathbf{r} = (\r^{(0)} \ \dots \ \r^{(N)}) \leftarrow \big(\extsmallfieldInverse(\R^{(0)}) \ \dots \ \extsmallfieldInverse(\R^{(N)}) \big)$\;
\BlankLine
$\c= (\c^{(0)} \ \c^{(1)} \ \dots \ \c^{(N)})\leftarrow$ \textsc{BoundedRowDistanceDecoderPUM}$\big(\r\big)$ with Algorithm~\ref{algo:pum}\;
\BlankLine
Reconstruct $\mathbf{u} = (\u^{(0)} \  \u^{(1)} \ \dots \u^{(N-1)})$\; 
\BlankLine
\KwOut{Information sequence $\u  = (\u^{(0)} \  \u^{(1)} \ \dots \u^{(N-1)}) \in \F^{kN} $}
 \vspace{1ex}}{0.52}

The asymptotic complexity of Algorithm~\ref{algo:dec_network} for decoding one matrix $\Y^{(i)}$ of size $n^{(i)} \times (n+m)$
scales cubic in $n$ over $\F$. Calculating the RRE is at most cubic in $n$ over $\Fq$ if we use Gaussian elimination. 
However, Algorithm~\ref{algo:pum} has asymptotic complexity $\mathcal O(n^3)$ over $\F$, which dominates therefore the complexity of Algorithm~\ref{algo:dec_network}. The reconstruction of the information sequence from the code sequence is negligible.

\begin{example}[Lifted PUM Code for Network Coding]\label{ex:decoding}
Let $N+1 = 7$, $n = 8 \leq m $, $k= 4$, $k^{(1)} = 2$ and therefore 
$d_0=d_1 = 5$, $d_{01}  = 7$ and $\da =  3$ (Table~\ref{tab:example}).
Let $\mycode{C}$ be a $\PUM{n,k}{k^{(1)}}$ code as in Definition~\ref{def:pumgab_genmat}.
Construct the lifting of $\mycode{C}$ as in Definition \ref{def:liftedpum}.

Assume, $\Y = (\Y^{(0)} \ \Y^{(1)} \ \dots \ \Y^{(6)})$ is given as output of the operator channel and apply Algorithm~\ref{algo:dec_network}.

After calculating the RRE (and filling the matrix with zero rows as in \eqref{eq:rre_network}), let the number of errors, row erasures and column erasures in each block be as in Table~\ref{tab:example}. The results of the different decoding steps of Algorithm~\ref{algo:pum} for error-erasure decoding of PUM codes are also shown. 
In this example the BRD condition \eqref{eq:bmdcond} is fulfilled and correct decoding is therefore guaranteed due to Theorem~\ref{thm:pum_decoding}.
 
The code rate of $\mycode{C}$ is $1/2$ and as a comparison with the (lifted) Gabidulin codes from \cite{silva_rank_metric_approach},  
the last line in Table~\ref{tab:example} shows the decoding of a block Gabidulin code of rate $1/2$ and minimum rank distance $d = 5$. For fairness, the last block is also decoded with a $\Gab{8,2}$ code.
The block decoder fails in Shots 1 and 5.

However, similar to the ongoing discussion whether block or convolutional codes are better, it depends on the distribution of the errors and erasures, i.e., on the channel, 
whether the construction from \cite{silva_rank_metric_approach} or ours performs better.
\end{example}

\begin{table*}[ht]
\caption{Example for error-erasure decoding of lifted (partial) unit memory codes based on Gabidulin codes.} 
\label{tab:example}
\centering
\vspace{1ex}
\begin{tabular}{p{0.12\textwidth} p{0.2\textwidth} p{0.03\textwidth}p{0.03\textwidth}p{0.03\textwidth}p{0.03\textwidth}p{0.03\textwidth}p{0.03\textwidth}p{0.02\textwidth} }
\toprule
&Shot $i$ & $0$ & $1$ & $2$ & $3$ & $4$ & $5$ & $6$ \\
\midrule
 &&&&&&&\\[-1.5ex]
&$\numbRowErasures^{(i)}+\numbColErasures^{(i)}$ &  $0$ & $1$ & $3$ & $1$ & $1$ & $0$ & $2$ \\
&$t^{(i)}$	&  $2$ & $2$ & $0$ & $1$ & $0$ & $3$ & $2$ \\
\midrule
 &&&&&&&\\[-1.5ex]
\textbf{PUM code} & Decoding with $\Ca$,\newline block $0$ with $\mycode{C}_0$,\newline block $N$ with $\mycode{C}_{10}$& \ \newline$\checkmark$ & $\times$ &$\times$ &$\times$ &$\checkmark$ &$\times$&\ \newline \ \newline$\checkmark$ \\
& Decoding with $\mycode{C}_0$, $\mycode{C}_1$ &  & $\times$ & $\checkmark$& $\checkmark$&  &$\times$ &\\
& Decoding with $\mycode{C}_{01}$ & & $\checkmark$ & & & & $\checkmark$ & \\[0.85ex]
\midrule
 &&&&&&&\\[-1.5ex]
\textbf{Block code} 
& Decoding with $\Gab{8,4}$& $\checkmark$ & $\times$ &$\checkmark$ &$\checkmark$ &$\checkmark$ &$\times$&$\checkmark$ \\
\bottomrule
\end{tabular}
\end{table*}

\section{Application to Random Affine\\ Network Coding}\label{sec:affine_networkcoding}
In this section, we outline the application of our construction of (P)UM codes in rank metric to error control in \emph{random affine network coding} (RANC), introduced by Gadouleau and Goupil in \cite{Gadouleau2011Matroid}. In this model, the transmitted packets are regarded as points in an affine space and the network performs \emph{affine} linear combinations of the received packets, i.e., the sum of the coefficients included in the linear combination equals one.
Instead of (linear) subspace codes, \emph{affine} subspace codes are considered, i.e., a code is a set of affine subspaces of an affine space, where an affine subspace of dimension $r$ is a linear subspace of dimension $r-1$, which is translated by one point.
RANC increases the data rate by around one symbol per packet compared to RLNC.  For details, the reader is referred to \cite{Gadouleau2011Matroid}.

Similar to the linear lifting of Definition~\ref{def:lifting_matrixcode}, an affine lifting can be used to construct affine subspace codes.
The affine lifting of a code $\mycode{C}$ is defined as follows:
Let $\widehat{\I}_{r-1} = [\0 \ \I_{r-1}]^T \in \Fq^{r \times (r-1)}$, where $\0=(0 \ 0 \ \dots 0)^T \in \Fq^{(r-1)\times 1}$. Then, the subspace $\liftmapaff{\X} = \Rowspace{[\widehat{\I}_{r-1} \ \X]}$ denotes the affine lifting of $\X \in \Fq^{r \times (n-r+1)}$. Compared to the linear lifting (Definition~\ref{def:lifting_matrixcode}), the overhead is reduced by one column and the size of $\X$ is increased by $r$ symbols, which makes affine lifting more efficient than linear lifting.

Based on the definition of affine lifting, we can immediately consider the affine lifting of our (P)UM code $\mycode{C}$ from Definition~\ref{def:liftedpum} by the following set of spaces:
\begin{align*}
\liftmapaff{\mycode{C}} \defeq \Big\lbrace \Big( \RowspaceNoInput \big( [& \widehat{\I}_{n-1}\  \C^{(0)T}]\big)  \ \dots \ \RowspaceNoInput \big( [\widehat{\I}_{n-1} \ \C^{(N)T}]\big) \Big) \\ &:\Big(\extsmallfieldInverse(\C^{(0)}) \ \dots \ \extsmallfieldInverse(\C^{(N)}) \Big) \in \mycode{C} \Big\rbrace,
\end{align*}
where $\widehat{\I}_{n-1} = [\0 \ \I_{n-1}]^T \in \Fq^{n \times (n-1)}$ and therefore the overhead is reduced by one column compared to Definition~\ref{def:liftedpum}. Alternatively, we can also define $\mycode{C}$ such that $\C^{(i)} \in \Fq^{(m+1) \times n}$, then the transmitted space has the same size as for RLNC, but we transmit $n$ additional information symbols over $\Fq$.

As shown in \cite[Section~VI]{Gadouleau2011Matroid}, the decoding of affine lifted codes is not more complicated than the one of linear lifted codes and, for our construction, it reduces in a similar way as in Section~\ref{sec:pum_networkcoding} to error-erasure decoding of the (P)UM code.

\section{Conclusion}\label{sec:conclusion}
In this paper, we have considered convolutional codes in rank metric, their decoding and their application to random linear network coding.

First, we have shown general distance measures for convolutional codes based on a modified rank metric---the sum rank metric---and have recalled upper bounds on the free rank distance and the slope of (P)UM codes based on the sum rank metric. 
Second, we have given an explicit construction of (P)UM codes based on the generator matrices of Gabidulin codes and have calculated its free rank distance and slope.
This (low-rate) construction achieves the upper bound on the free rank distance. We have also generalized this construction to arbitrary code rates.
Third, we have presented an efficient error-erasure decoding algorithm for our (P)UM construction.
The algorithm guarantees to correct errors up to half the active row rank distance and its complexity is cubic in the length.
Finally, we have shown how constant-dimension codes, which were constructed by lifting the (P)UM code, can be applied for error control in random linear network coding and outlined the application of (P)UM codes in rank metric to affine linear network coding.

\vspace{-1ex}

\section*{Acknowledgment}
The authors would like to thank Martin Bossert, Alexander Zeh, and Victor Zyablov for the valuable discussions and the reviewers for their very helpful comments.

\bibliographystyle{IEEEtran}
\bibliography{antoniawachter}

\end{document}